\documentclass[sigconf]{acmart}

\AtBeginDocument{%
  \providecommand\BibTeX{{%
    \normalfont B\kern-0.5em{\scshape i\kern-0.25em b}\kern-0.8em\TeX}}}

\copyrightyear{2022} 
\acmYear{2022} 
\setcopyright{acmcopyright}\acmConference[ICSE '22]{44th International Conference on Software Engineering}{May 21--29, 2022}{Pittsburgh, PA, USA}
\acmBooktitle{44th International Conference on Software Engineering (ICSE '22), May 21--29, 2022, Pittsburgh, PA, USA}
\acmPrice{15.00}
\acmDOI{10.1145/3510003.3510050}
\acmISBN{978-1-4503-9221-1/22/05}

\usepackage{graphicx}
\usepackage{amsmath}
\usepackage[american]{babel}
\usepackage{microtype}
\usepackage{subcaption}
\usepackage[ruled,,linesnumbered ]{algorithm2e}
\usepackage{tcolorbox}
\usepackage{balance} 
\usepackage{enumitem} 
\usepackage{color}

\newcount\DraftStatus  
\definecolor{darkgreen}{rgb}{0,0.5,0} 
\definecolor{purple}{rgb}{1,0,1} 
\definecolor{todocolor}{rgb}{0.9,0.1,0.1} 
\definecolor{fixcolor}{rgb}{0.1,0.7,0.3} 
\definecolor{wycolor}{rgb}{0.9,0.1,0.1} 
\definecolor{hycolor}{rgb}{0.7,0.7,0.3} 

\newcommand{\nbc}[3]{\ifnum\DraftStatus=1
	{\colorbox{#3}{\bfseries\sffamily\scriptsize\textcolor{white}{#1}}}
	{\textcolor{#3}{\sf\small$\blacktriangleright$\emph{#2}$\blacktriangleleft$}}
\fi}

\newcommand{\draftnote}[2]{\ifnum\DraftStatus=1
	\marginpar{
		\tiny\raggedright
		\hbadness=10000
		\def\baselinestretch{0.8}
		\textcolor{#1}{\textsf{\hspace{0pt}#2}}}
\fi}

\begin{document}
\title{What Do They Capture? - A Structural Analysis of Pre-Trained Language Models for Source Code}

\author{Yao Wan}
\authornote{Also with National Engineering Research Center for Big Data Technology and System, Services Computing Technology and System Lab, Cluster and Grid Computing Lab, HUST, Wuhan, 430074, China.}
\affiliation{%
 \institution{School of Computer Science and Technology, Huazhong University of Science and Technology, China}
 \streetaddress{Luoyu Rd 1037}
 \country{}
 }
\email{wanyao@hust.edu.cn}

\author{Wei Zhao}
\authornotemark[1]
\affiliation{%
 \institution{School of Computer Science and Technology, Huazhong University of Science and Technology, China}
 \country{}
 }
\email{mzhaowei@hust.edu.cn}

\author{Hongyu Zhang}
\affiliation{%
\institution{University of Newcastle}
\country{Australia}
}
\email{hongyu.zhang@newcastle.edu.au}

\author{Yulei Sui}
\affiliation{%
\institution{School of Computer Science, University of Technology Sydney}
\country{Australia}
}
\email{yulei.sui@uts.edu.au}

\author{Guandong Xu}
\affiliation{%
\institution{School of Computer Science, University of Technology Sydney}
\country{Australia}
}
\email{guandong.xu@uts.edu.au}

\author{Hai Jin}
\authornotemark[1]
\affiliation{%
 \institution{School of Computer Science and Technology, Huazhong University of Science and Technology, China}
 \country{}
 }
\email{hjin@hust.edu.cn}
%

\begin{abstract}
Recently, many pre-trained language models for source code have been proposed to model the context of code and serve as a basis for downstream code intelligence tasks such as code completion, code search, and code summarization.
These models leverage masked pre-training and Transformer and have achieved promising results. However, currently there is still little progress regarding  interpretability of existing pre-trained code models. 
It is not clear \textit{why} these models work and \textit{what} feature correlations they can capture.
In this paper, we conduct a thorough structural analysis aiming to provide an interpretation of pre-trained language models for source code (e.g., CodeBERT, and GraphCodeBERT) from three distinctive perspectives: (1) attention analysis, (2) probing on the word embedding, and (3) syntax tree induction.
Through comprehensive analysis, this paper reveals several insightful findings that may inspire future studies: (1) Attention aligns strongly with the syntax structure of code. (2) Pre-training language models of code can preserve the syntax structure of code in the intermediate representations of each Transformer layer. (3) The pre-trained models of code have the ability of inducing syntax trees of code. Theses findings suggest that it may be helpful to incorporate the syntax structure of code into the process of pre-training for better code representations.
\end{abstract}

%
%

%

\maketitle
\section{Introduction}
    Code representation learning (also known as code embedding) aims to encode the code semantics into distributed vector representations, and plays an important role in recent deep-learning-based models for code intelligence. Code embedding can be used to support a variety of downstream tasks, such as code completion~\cite{raychev2014code}, code search~\cite{gu2018deep,wan2019multi}, and code summarization~\cite{alon2018code2seq,wan2018improving,zhangretrieval20}.

    Current approaches to code embedding mainly fall into two categories from the perspectives of supervised and unsupervised (or self-supervised) learning paradigms.
    The supervised approaches are typically developed for  specific tasks following the \textit{encoder-decoder} architecture~\cite{sutskever2014sequence}. In this architecture, an \textit{encoder} network (e.g. LSTM, CNN, and Transformer) is used to produce a vector representation of a program. The resulting vector is then fed as an input into a \textit{decoder} network  to perform some prediction tasks, e.g., summary generation~\cite{alon2018code2seq,wan2018improving,sui2020flow2vec} or token sequence  prediction~\cite{raychev2014code}.
    Recently, there has been significant improvement in the expressiveness of models that can learn the semantics of code, such as self-attention based architectures like Transformer~\cite{vaswani2017attention}.
    Another line of code embedding research is based on unsupervised learning.
    Some approaches utilize word embedding techniques to represent source code~\cite{mikolov2013efficient,mikolov2013distributed}, which aim to learn a global word embedding matrix $\mathbf{E}\in\mathbb{R}^{V\times D}$, where $V$ is the vocabulary size and $d$ is the number of dimensions.
    Code2Vec~\cite{alon2019code2vec} is such kind of approach, which learns a distributed representation of code based on the sampled paths from \textit{Abstract Syntax Trees} (ASTs). 
    
    Recently, self-supervised models which are pre-trained through masked language modeling have attracted much attention. Pre-trained models such as BERT~\cite{devlin2019bert} and ELMo~\cite{peters2018deep} are representative approaches and have been successfully used in a variety of tasks in NLP.
    Inspired by the self-supervised pre-training in NLP, there have been some recent efforts in developing pre-trained models on large-scale code corpus for software engineering tasks.
    For example, CuBERT~\cite{kanade2020learning} is a pre-trained BERT model using 7.4M Python files from GitHub.
    CodeBERT~\cite{feng2020codebert} is a bimodal pre-trained model on 
    source code and natural-language descriptions.
    To incorporate the syntax structure of code, \citet{guo2020graphcodebert} further propose GraphCodeBERT to preserve the syntax structure of source code by introducing an  edge masking technique over data-flow graphs.
    With much effort being devoted to pre-trained code embedding, there is a pressing demand to understand \textit{why} they work and \textit{what} feature corrections they are capturing.
    
    In the NLP community, several recent studies have been made towards interpreting the pre-trained language models, e.g., BERT, from the perspective of attention analysis and task probing.
    This kind of research has become a subspecialty of ``BERTology''~\cite{rogers2020primer}, which focuses on studying the inner-mechanism of BERT model~\cite{devlin2019bert}. 
    However, in software engineering, such an understanding is yet to be achieved. Often, we see pre-trained language models 
    that achieve superior performance in various software engineering tasks, but do not understand why they work. 
    Currently there have been several empirical studies that aim to investigate the effectiveness of code embedding. For example, \citet{kang2019assessing} empirically assessed the impact of code embedding for different downstream tasks.
    \citet{chirkova2020empirical} conducted another study to investigate the capabilities of Transformers to utilize syntactic information in different tasks. However, these studies only show in which scenarios a code embedding technique works better, without explaining the inner-mechanism of why the embedding achieves good results.
    Therefore, it is still not clear why the pre-trained language models work and what they indeed capture, in the context of software engineering tasks.     

    In this work, we explore the interpretability of pre-trained code models.  
    More specifically, we try to answer the following question: \textit{Can the existing pre-trained language models learn the syntactical structure of source code written in programming languages?}
    Addressing this question plays an important role in understanding  the learned structure of deep neural networks. 
    We conduct a thorough structural analysis from the following three aspects, aiming to provide an interpretation of pre-trained code models (e.g., CodeBERT, GraphCodeBERT).
    \vspace{-6pt}
    \begin{itemize}
    \item As the first contribution, we analyze the self-attention weights and align the weights with the syntax structure (see Sec.~\ref{sec_attention_analysis}).
    Given a code snippet, our assumption is that if two tokens are close to each other in the AST, i.e., have a neighbourhood relationship, the attention weights assigned to them should be high. 
    Our analysis reveals that the attention can capture high-level structural properties of source code, i.e., the \textit{motif structure} in ASTs.
    
    \item As the second contribution, we design a structural probing approach ~\cite{hewitt2019structural} to investigate whether the syntax structure is embedded in the linear-transformed contextual word embedding of pre-trained code models (see Sec.~\ref{sec_structural_probing}). 
    Using our probe, we show that such transformations also exist in the pre-trained language models for source code, showing evidence that the syntax structure of code is also 
    embedded implicitly in the vectors learned by the model.
     \item As the third contribution, we investigate whether the pre-trained language models for source code provide the ability of inducing the syntax tree without training (see Sec.~\ref{sec_grammar_induction}). We find that the pre-trained models can indeed learn the syntax structure of source code to a certain extend.
       \end{itemize}
    
    Our work is complementary to other works that aim to design better neural networks for source code representation.
    We believe that the findings revealed in this paper may shed light on the inner mechanism of pre-training models for  programming languages, as well as inspire further studies.

\section{Background}\label{sec_background}
\begin{figure}[t!]
	\centering
	\includegraphics[width=.4\textwidth]{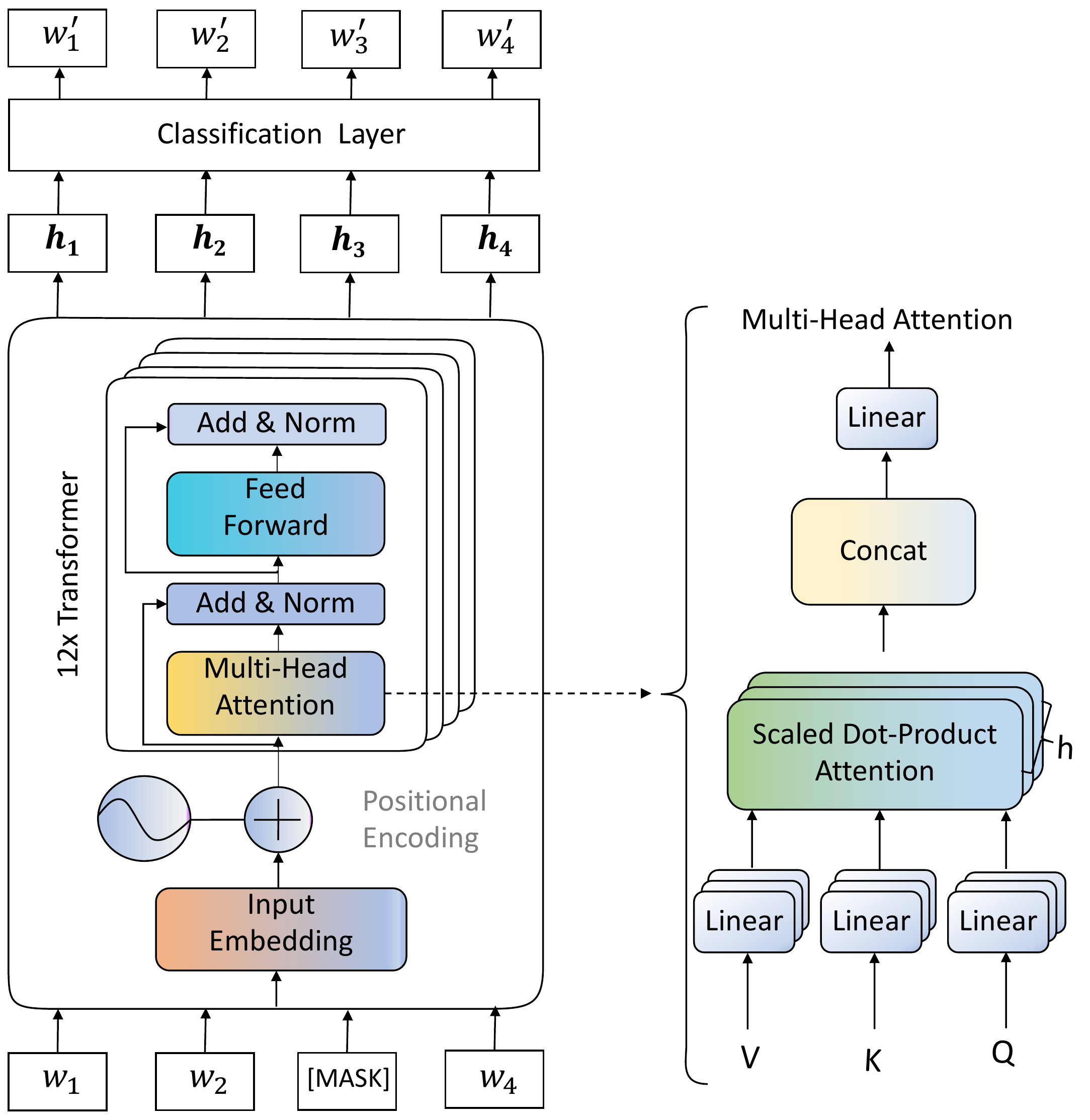}
	\caption{A general framework for Transformer-based language model pre-training~\cite{devlin2019bert}.}
	\label{fig_bert}
	\vspace{-1em}
\end{figure}

In this section, we introduce some background knowledge of our work, including Transformer, and pre-training language model.
Figure~\ref{fig_bert} shows a general framework for Transformer-based language model pre-training.

\begin{figure*}[t]
	\begin{subfigure}[b][][c]{.28\textwidth}
		\centering
		\includegraphics[width=\textwidth]{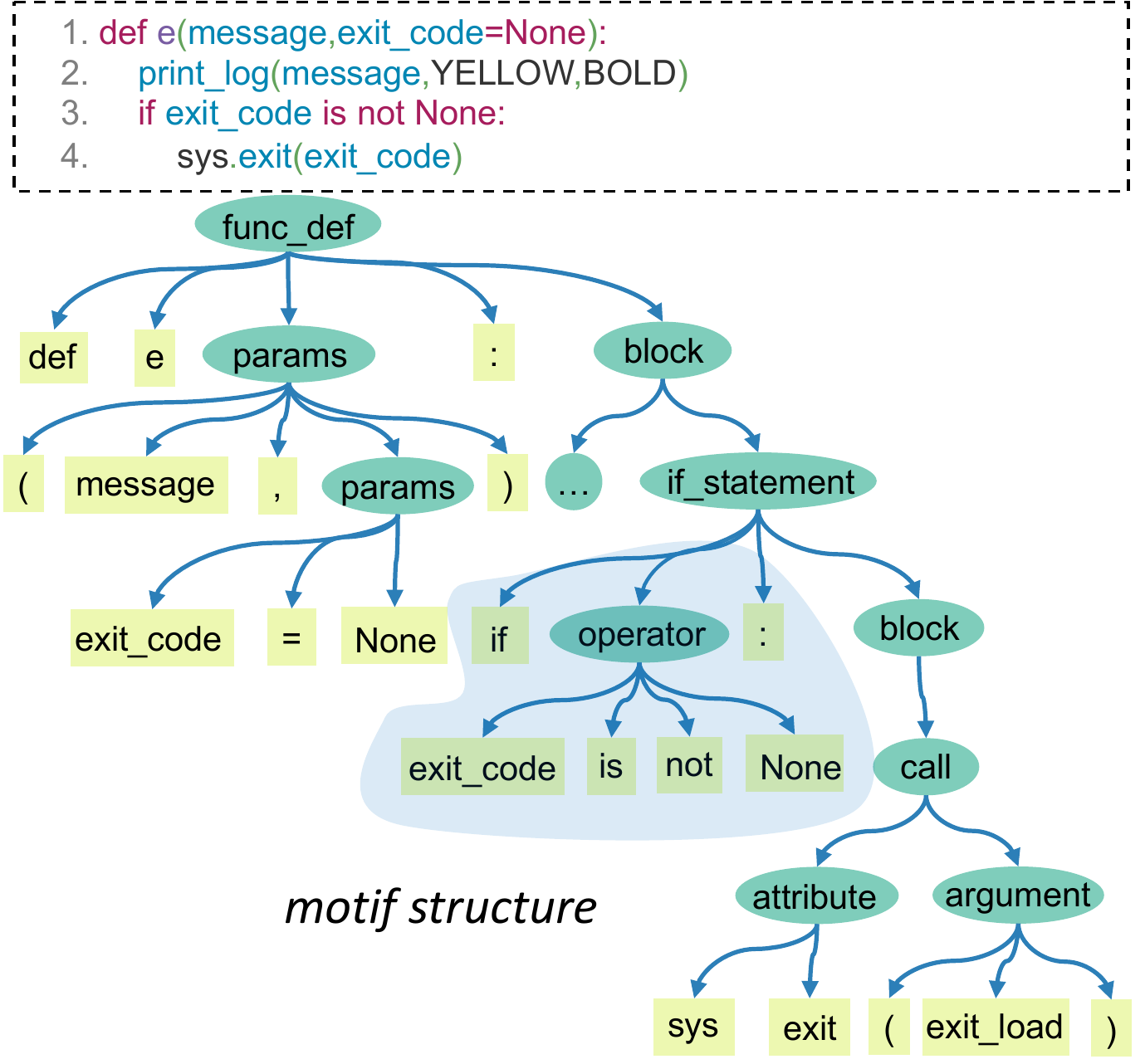}
		\caption{A Python code snippet with its AST}
		\label{fig:figure1-a}
	\end{subfigure}
	\hspace{1em}
	\begin{subfigure}[b][][c]{.35\textwidth}
		\centering
		\includegraphics[width=\textwidth]{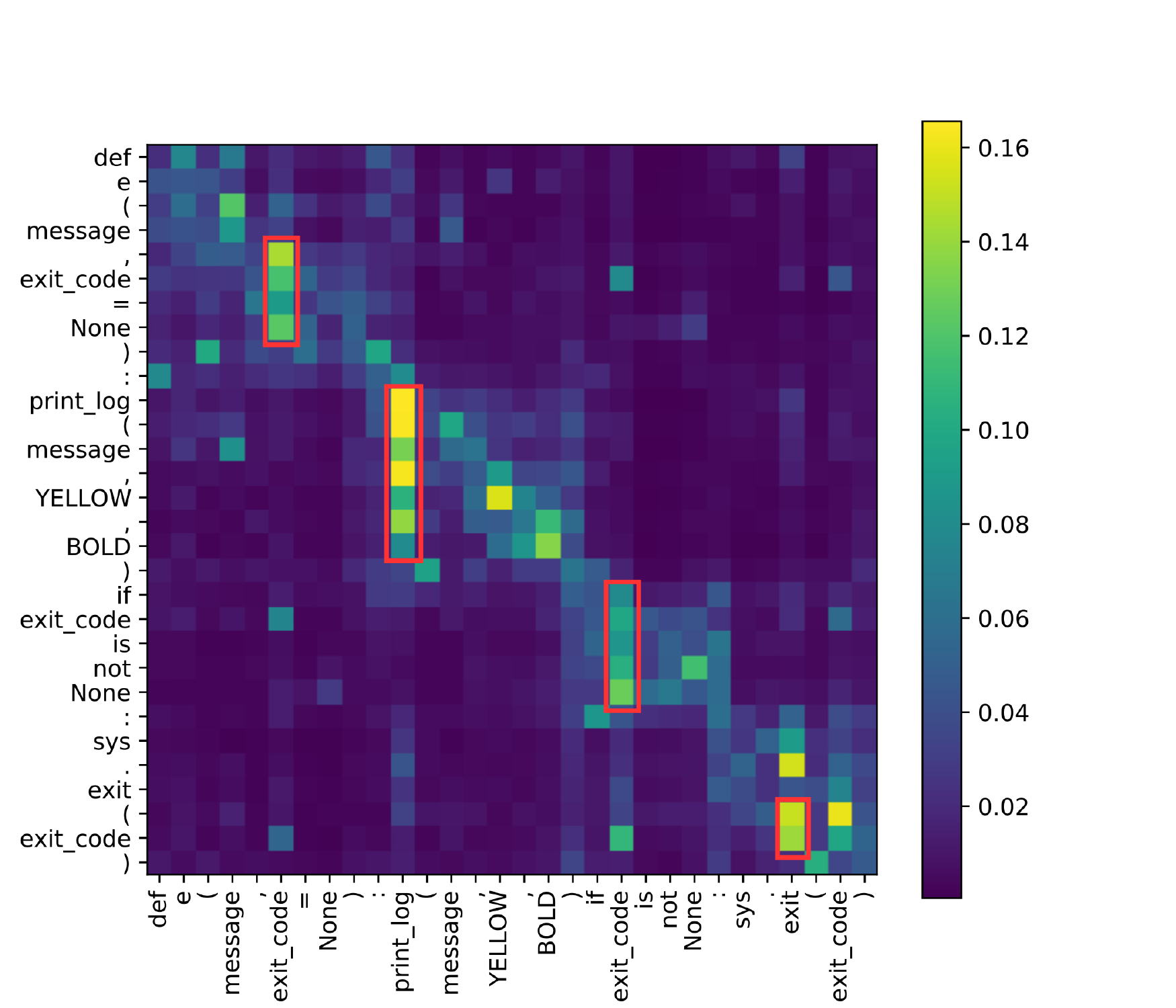}
		\caption{Attention heatmap in Layer 5}
		\label{fig:figure1-b}
	\end{subfigure}
	\begin{subfigure}[b][][c]{.32\textwidth}
		\centering
		\includegraphics[width=\textwidth]{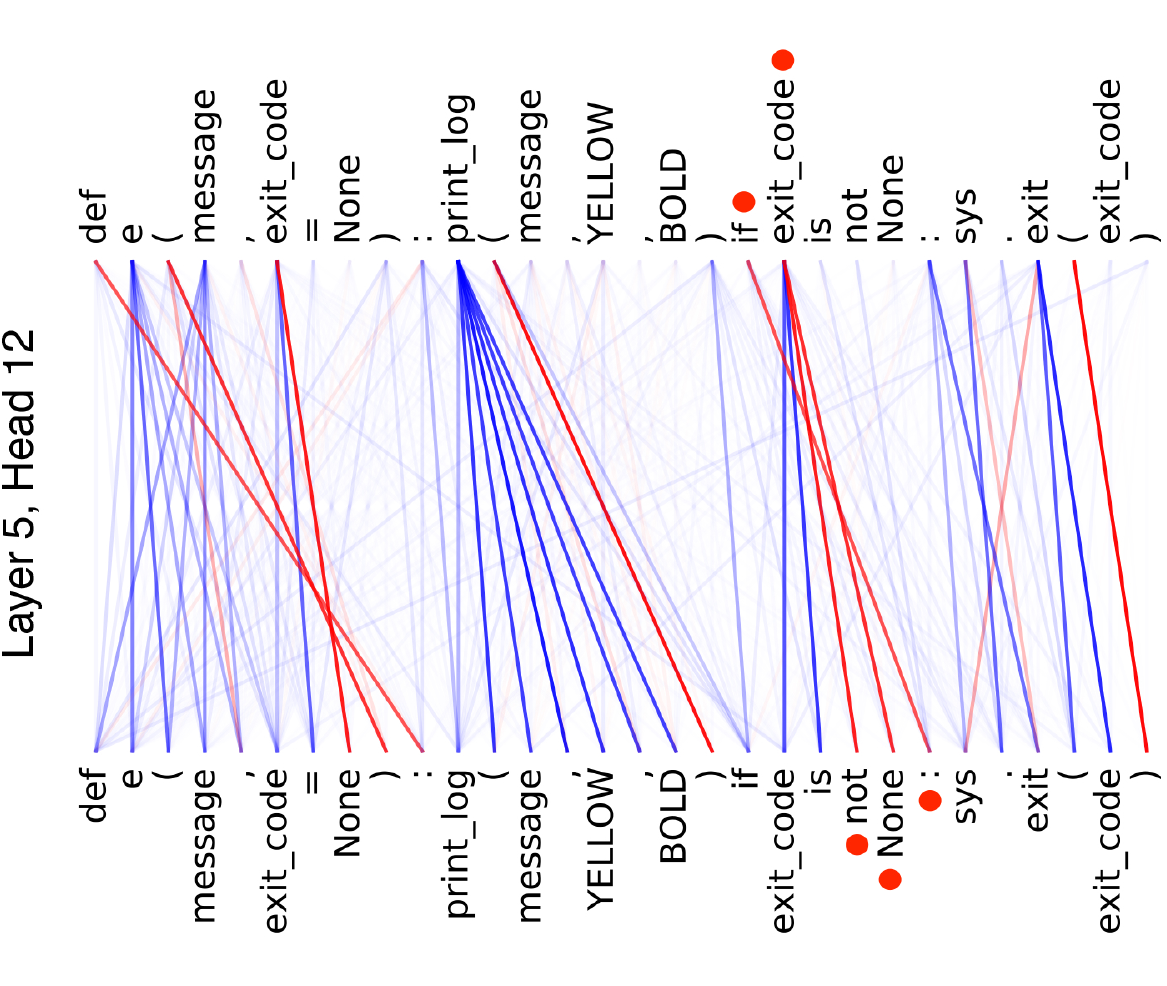}
		\caption{Attention distribution in Layer 5, Head 12}
		\label{fig:figure1-c}
	\end{subfigure}
	\caption{
	Visualization of self-attention distribution for a code snippet in CodeBERT. (a) A Python code snippet with its corresponding AST. (b) Heatmap of the averaged attention weights in Layer 5. (c) Self-attention distribution in Layer 5, Head 12. 
    The brightness of lines indicates the attention weights in a specific head. If the connected nodes appear in the \textit{motif structure} of the corresponding AST, we mark the lines in red.
	}
	\label{fig_motivation}
	\vspace{-1em}
\end{figure*}
\subsection{Self-Attention-Based Transformer}
Transformer~\cite{vaswani2017attention}, which is solely based on self-attention, has become a popular component for code representation learning.
Let $c=\{w_1, w_2,\ldots,w_n\}$ denote a code snippet of a sequence of tokens with length of $n$.
A Transformer model is composed of $L$ layers of Transformer blocks to represent a code snippet into contextual representation at different levels  $\mathbf{H}^{l}=[ \mathbf{h}_1^l, \mathbf{h}_2^l,\ldots,\mathbf{h}_{n}^l ]$, where $l$ denotes the $l$-th layer. 
For each layer, the layer representation $\mathbf{H}^{l}$ is computed by the $l$-th layer Transformer block $\mathbf{H}^{l} = \mathrm{ Transformer}_{l}(\mathbf{H}^{l-1})$, $l \in \{1,2,\ldots,L\}$.

In each Transformer block, multiple self-attention heads are used to aggregate the output vectors of the previous layer. 
Given an input~$c$, the self-attention mechanism assigns each token $w_i$ a set of attention weights over the token in the input:
\begin{equation}
    \operatorname{Attn}(w_i) = (\alpha_{i,1}(c), \alpha_{i,2}(c),\ldots,\alpha_{i,n}(c))\,,
\end{equation}
where $\alpha_{i,j}(c)$ is the attention that $w_i$ pays to $w_j$.
The attention weights are computed from the scaled dot-product of the \textit{query vector} of $w_i$, and the \textit{key vector} of $w_j$, followed by a softmax.
In the vectorized computing, a general attention mechanism can be formulated as the weighted sum of the value vector $\mathbf{V}$, using the query vector $\mathbf{Q}$ and the key vector $\mathbf{K}$:
\begin{equation}
	\operatorname{Att}(\mathbf{Q}, \mathbf{K}, \mathbf{V})=\operatorname{softmax}\left(\frac{\mathbf{Q} \mathbf{K}^{T}}{\sqrt{d_{\text {model}}}}\right) \cdot \mathbf{V}\,,
\end{equation}
where $d_{\rm model}$ represents the dimension of each hidden representation. For self-attention, $\mathbf{Q}$, $\mathbf{K}$, and $\mathbf{V}$ are mappings of the previous hidden representation by different linear functions, i.e., 
$\mathbf{Q} =\mathbf{H}^{l-1} \mathbf{W}_{Q}^{l}$, $\mathbf{K} =\mathbf{H}^{l-1} \mathbf{W}_{K}^{l}$, and  $\mathbf{V} =\mathbf{H}^{l-1} \mathbf{W}_{V}^{l}$, respectively. At last, the encoder produces the final contextual representation $\mathbf{H}^{L} = [\mathbf{h}^{L}_{1}, \ldots, \mathbf{h}^{L}_{n}]$, which is obtained from the last Transformer block.

In order to utilize the order of the sequential tokens, the ``positional encodings'' are injected to the input embedding.
\begin{equation}
	\mathbf{w}_i = e(w_i) +  pos(w_i)\,,
\end{equation}
where $e$ denotes the word embedding layer, and $pos$ denotes the positional embedding layer. Typically, the positional encoding implies the position of code token based on sine and cosine functions.


\subsection{Pre-Training Language Model}
\begin{figure*}[t!]
	\centering
	\includegraphics[width=\textwidth]{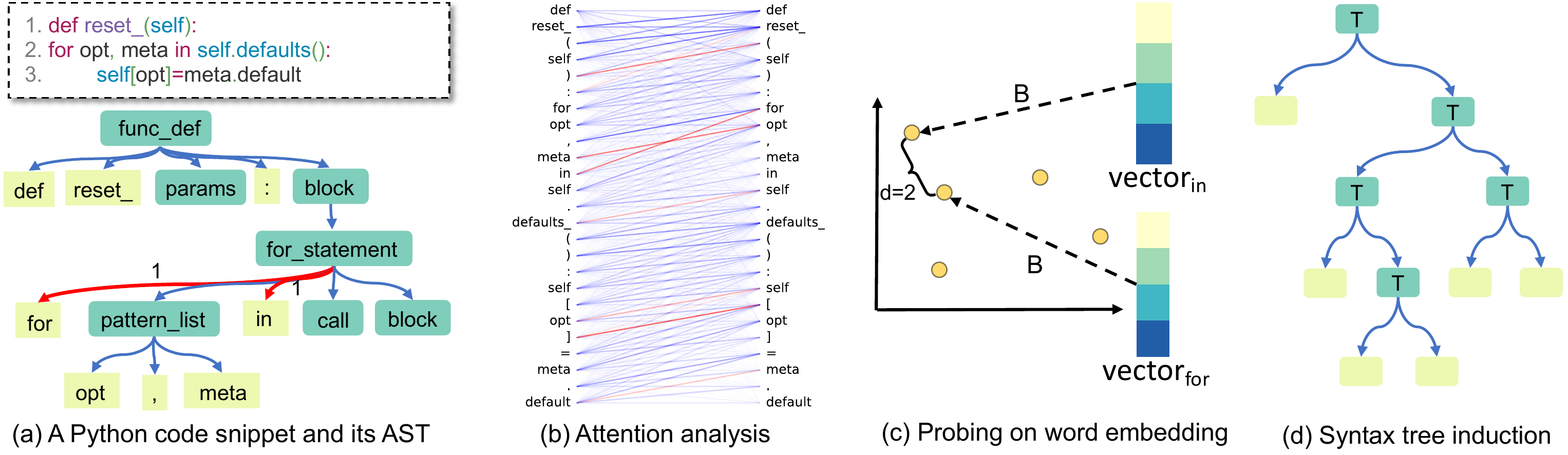}
	\caption{
	An illustration of attention analysis, probing on word embedding, and syntax tree induction, with a specific Python code snippet.
	}
	\label{fig_methods_illustration}
	\vspace{-1em}
\end{figure*} 
Given a corpus, each sentence (or code snippet) is first  tokenized into a series of tokens (e.g., \textit{Byte Pair Encoding}, BPE~\cite{sennrich2015neural}).
Before BERT's pre-training, it takes the concatenation of two segments as the input, defined as $c_1=\{w_1, w_2,\ldots,w_n\}$ and $c_2=\{u_1, u_2, \ldots, u_m\}$, where $n$ and $m$ denote the lengths of two segments, respectively. 
The two segments are always connected by a special separator token \texttt{[SEP]}.
The first and last tokens of each sequence are always padded with a special classification token \texttt{[CLS]} and an ending token \texttt{[EOS]}, respectively.
Finally, the input of each training sample will be represented as follows: 
$$
s=\texttt{[CLS]}, \underset{c_1}{\underbrace{w_1, w_2, \ldots, w_n}}, \texttt{[SEP]}, \underset{c_2}{\underbrace{u_1, u_2, \ldots, u_m}}, \texttt{[EOS]}\,.
$$

The input is then fed into a Transformer encoder.
During BERT's pre-training, two objectives are designed for self-supervised learning, i.e., \textit{masked language modeling} (MLM) and \textit{next sentence prediction} (NSP).
In the masked language modeling, the tokens of an input sentence are randomly sampled and replaced with the special token \texttt{[MASK]}.
In practice, BERT uniformly selects 15\% of the input tokens for possible replacement. Among the selected tokens, 80\% are replaced with \texttt{[MASK]}, 10\% are unchanged, and the left 10\% are randomly replaced with the selected tokens from vocabulary~\cite{devlin2019bert}.
%
For next sentence prediction, it is modeled as a binary classification to predict whether two segments are consecutive. 
Training positive and negative examples are conducted based on the following rules: (1) if two sentences are consecutive, it will be considered as a positive example; (2) otherwise, those paired segments from different documents are considered as negative examples. 
Recently, self-supervised learning using masked language modeling has become a popular technique for natural language understanding and generation ~\cite{devlin2019bert,liu2019roberta,Clark2020ELECTRA,sun2019ernie,song2019mass,dong2019unified}.
In the context of software engineering, several pre-trained code models have also been proposed for program understanding. 
In this paper, we select two representative pre-trained models for code representations:
(1) CodeBERT~\cite{feng2020codebert}, which takes the concatenation of source code and natural-language description as inputs, and pre-trains a language model by masking the inputs;
and (2) GraphCodeBERT~\cite{guo2020graphcodebert}, which improves CodeBERT by incorporating the data-flow information among variables into model pre-training.

\section{Motivation}\label{sec_motivating_example}
Prior work in NLP has pointed out that the self-attention mechanism in Transformer has the capability of capturing certain syntax information in natural languages.
Inspired by this, we visualize and investigate the attention distribution of the pre-trained model (i.e., CodeBERT) for a code snippet, as shown in Figure~\ref{fig_motivation}. 
Figure~\ref{fig_motivation}(a) shows a Python code snippet with its AST.
In this paper, we define the syntax structure of the AST consisting of a non-leaf node with its children (e.g., \texttt{if\_statement} and \texttt{block} in Figure~\ref{fig_motivation}(a)) as a \textit{motif structure}.
We believe that the syntax information of code can be represented by a series of \textit{motif structures}.

Given a code snippet and its corresponding AST, Figure~\ref{fig_motivation}(b) visualizes the self-attention heatmap for a specific layer (i.e., Layer 5), which is an average of attention weights over multiple heads.
From this figure, we can observe that several patterns indeed exist in the self-attention heatmap, depicted as groups of rectangles (marked in red). These rectangles indicate that the code tokens form groups.
Interestingly, we can also find that each group of tokens is close to each other in the AST. 
Taking ``\texttt{if exit\_code is not None}'' as an example, which is an \texttt{if} statement, 
we can see that, in the AST, all of these tokens are in the same branch of \texttt{if\_statement}.
In addition, we can see that these code tokens are also closely connected in the self-attention heatmap.

Moreover, we also visualize the self-attention distribution in a specific head (Layer 5, Head 12) to analyze the connections between two tokens, as shown in Figure~\ref{fig_motivation}(c).
In this figure, the brightness of lines indicates the attention weights in a specific head. If the connected nodes appear in the \textit{motif structure} of the corresponding AST, we mark the lines in red.
From this figure, we can observe that those code tokens (i.e., ``\texttt{if}'', ``\texttt{exit\_code}'', ``\texttt{not}'', and ``\texttt{None}'') that are in a \textit{motif structure} indeed have been highlighted as closely connected by self-attention.

As we have identified several patterns from the attention distribution, which provide some hints to the syntax structure of code, 
it is necessary for us to further explore this phenomenon with quantitative analysis and systematic assessment.
Motivated by the aforementioned observations, this paper investigates \textit{why} pre-trained language models for source code work and \textit{what} feature correlations they are capturing, by analyzing the self-attention mechanism.
In particular, we analyze two outputs of the self-attention mechanism, i.e., the attention distribution and the generated hidden vectors, under the framework of Transformer.

\section{Structural Analysis of Pre-Trained Language Models for Source Code}\label{sec_structural_analysis}
In this section, we propose three distinct structural analysis approaches, i.e., attention analysis, structural probing on word embedding, and syntax tree induction, to interpret pre-trained code models (i.e., CodeBERT and GraphCodeBERT). 
Figure~\ref{fig_methods_illustration} gives an illustration of the three structural analysis approaches.
Before introducing each approach, we first introduce several common notations that will be used later.
Let $(w_1,w_2,\ldots,w_n)$ denote the code token sequence of code snippet $c$, with length $n$. 
On the $l$-th layer of Transformer, we use $(\mathbf{h}_1^l,\mathbf{h}_2^l,\ldots,\mathbf{h}_n^l)$ to denote the sequence of contextual representation of each code token. 

\subsection{Attention Analysis}\label{sec_attention_analysis}
We start by analyzing the self-attention weights, which are the core mechanism for pre-training Transformer-based models.
Intuitively, attention defines the closeness of each pair of code tokens.
From the lens of attention analysis, we aim to analyze how attention aligns with the syntactical relations in source code.
In particular, we consider the syntactical relations such that the attention weight is high between two AST tokens sharing the same parent node. 
Figures~\ref{fig_methods_illustration}(a) and \ref{fig_methods_illustration}(b) illustrate the attention analysis. Given a code snippet with its AST, we can see that the leaf nodes \texttt{for} and \texttt{in} share the same parent. 
As expected, this structure is aligned with the attention weight $\alpha_{\texttt{for}, \texttt{in}}$ between these two nodes. 

Specifically, on each Transformer layer, we can obtain a set of attention weights $\alpha$ over the input code, where $\alpha_{i,j}>0$ is the attention from $i$-th code token to $j$-th token.
Here, we define an indicator function $f(w_i,w_j)$ that returns $1$ if $w_i$ and $w_j$ are in a syntactic relation ($w_i$ and $w_j$ have the same parent node in the AST)\footnote{Note that, in this paper, we exclude the condition that $w_i$ and $w_j$ are adjacent in textual appearance.}, and $0$ otherwise. 
We define the attention weight between $w_i$ and $w_j$ as $\alpha_{i,j}(c)$, and if $w_i$ and $w_j$ are very close, the attention weight should be larger than a threshold, i.e., $\alpha_{i,j}(c)>\theta$.
Therefore, the proportion of high-attention token pairs ($\alpha_{i,j}(c)>\theta$) aggregated over a dataset $\mathcal{C}$ can be formulated as follows: 
\begin{equation}
    p_\alpha (f)=\frac{\sum\limits_{c\in \mathcal{C}} \sum\limits_{i=1}^{\left |c  \right |}\sum\limits_{j=1}^{\left | c \right |} \vmathbb{1}_{\alpha_{i,j}(c)> \theta} \cdot f(w_i,w_j)  }{\sum\limits_{c\in \mathcal{C}} \sum\limits_{i=1}^{\left |c  \right |}\sum\limits_{j=1}^{\left | c \right |} \vmathbb{1}_{\alpha_{i,j}(c)> \theta}}\,,
    \label{eq_attention_analysis}
\end{equation}
where $\theta$ is a threshold selected for high-confidence attention weights.

\paragraph{Variability}
Equation~\ref{eq_attention_analysis} shows that, the portion of aligned attention is only dependent on the absolute value of attention weight $\alpha_{i,j}(c)$.
We hypothesize that those heads
that are attending to the position,
i.e., those heads that focus on the previous or next code token, would not align well with the syntax structure of code, since they do not consider the content of the code token. 
To distinguish whether the heads are attending to content or position of the code token, 
we further investigate the attention variability, which measures how attention varies over different inputs. The attention variability is formally defined as follows~\cite{vig2019analyzing}:
\begin{equation}\label{eq_variability}
Variability_\alpha=\frac{\sum\limits_{c\in C} \sum\limits_{i=1}^{\left |c  \right |}\sum\limits_{j=1}^{\left | i \right |}\left | \alpha_{i,j}(c)-\bar{\alpha }_{i,j}  \right |  }{2\cdot\sum\limits_{c\in C} \sum\limits_{i=1}^{\left |c  \right |}\sum\limits_{j=1}^{\left | i \right |} \alpha_{i,j}(c)}\,,
\end{equation}
where $\bar{\alpha}_{i,j}$ is the mean of $\bar{\alpha}_{i,j}(c)$ over all $c\in \mathcal{C}$. We only include the first $N$ tokens ($N=10$) of each $c\in \mathcal{C}$ to ensure a sufficient amount of data at each position $i$. The positional patterns appear to be consistent across the entire sequence.
The high variability would suggest a content-dependent head, while low variability would indicate a content-independent head.

\begin{figure}[!ht]
	\centering
	\includegraphics[width=.24\textwidth]{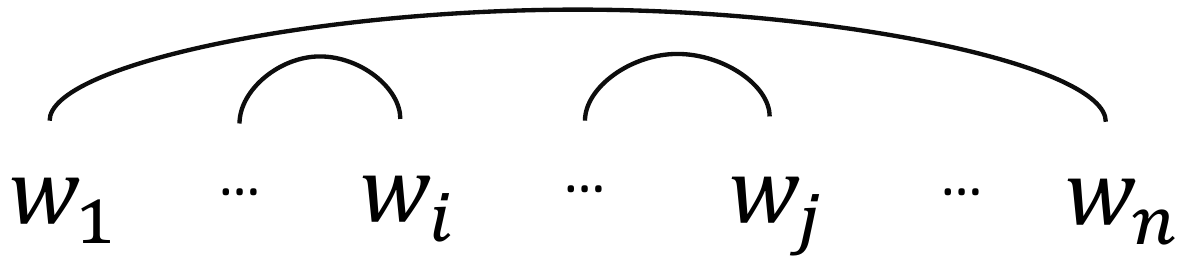}
	\caption{An illustration of the connection between distance and the syntax structure.}
	\label{fig_distance_vs_syntax}
	\vspace{-1em}
\end{figure} 
\subsection{Structural Probing on Word Embedding}\label{sec_structural_probing}
In this approach, we propose a structural probing analysis approach to investigate whether a pre-trained model embeds the syntactical structure in its contextual word embedding.
The key idea of our approach is that tree structure is embedded if the transformed space has the property that the Euclidean distance between two words' vectors corresponds to the number of edges between the words in the syntax tree.
One question may arise: \textit{why does the distance between nodes in the syntax tree matter for syntax information.}
This is because the distance metric (i.e., the path length between each pair of words) can recover the syntax tree simply by identifying that nodes $u$ and $v$ are neighbors if the distance between them equals to~1.
This has also been shown in the Code2Vec~\cite{alon2019code2vec}, which utilizes the contextual information among a set of paths sampled from the AST, to represent the structure information of code.
Figure~\ref{fig_distance_vs_syntax} gives a toy example to illustrate the connection between distance and syntax structure. 
Let $(w_1,\ldots,w_i,\ldots,w_j,\ldots,w_n)$ denote the sequence of code tokens for code snippet $c$, if we know the distance between every pair of nodes, we can induce the syntax structure of code.
Note that, the distance metric, which measures the distance among any two code tokens, can learn the global syntax structure of code to some extent.

Figures~\ref{fig_methods_illustration}(a) and \ref{fig_methods_illustration}(c) illustrate the structural probing on word embedding. 
Taking the leaf nodes \texttt{for} and \texttt{in} which share the same parent as an example, the Euclidean square of distance between these two nodes is 2. 
We first map the representations of these two tokens into a hidden space via a linear transformation $B$, obtaining the $Vector_{\texttt{for}}$, and $Vector_{\texttt{in}}$, respectively.
We believe that if the Euclidean square of distance between $Vector_{\texttt{for}}$, and $Vector_{\texttt{in}}$ is close to 2, the syntax structure between \texttt{for} and \texttt{in} is well preserved.

In particular, we learn the mapping function $B$ in a supervised way.
Formally, given a code sequence $(w_1,w_2,\ldots,w_n)$ as the input, and each model layer generates word vectors $(\mathbf{h}_1,\mathbf{h}_2,\ldots,\mathbf{h}_n)$. We compute the square of distance between the two word vectors $\mathbf{h}_i$ and $\mathbf{h}_j$ in high-dimensional hidden space as follows:
\begin{equation}\label{eq_mapping_matrix}
    d_B(\mathbf{h}_{i},\mathbf{h}_{j})^2=(B(\mathbf{h}_{i}-\mathbf{h}_{j}))^T(B(\mathbf{h}_{i}-\mathbf{h}_{j}))\,,
\end{equation}
where $i$ and $j$ are indices of the words in the code sequence. 
The parameters of the structural probe we use are exactly the matrix $B$ (liner mapping), which is trained to reconstruct the tree distance between all word pairs ($w_i,w_j$) in the code sequence $T$ in the training set of source code. 
We define the loss function for parameter training as follows:
\begin{equation}\label{eq_mapping_matrix_learning}
    \min_{B}\sum_{c\in\mathcal{C}}\frac{1}{\left | c \right |^2}\sum\limits_{i,j}\left | d_{T^c}(w_{i}^{c},w_{j}^{c})- d_B(\mathbf{h}_{i}^{c},\mathbf{h}_{j}^{c})^2\right |\,,
\end{equation}
where $ \left | c \right |$ is the length of code sequence $c$, $d_{T^c}(w_{i}^{c},w_{j}^{c})$ denotes the distance between code tokens in AST,
and $d_B(\mathbf{h}_{i}^{c},\mathbf{h}_{j}^{c})^2$ denotes the distance between the embedding vectors of code tokens, for code sequence $c$. 
The first summation calculates the average distance for all training sequences, while the second one sums all possible combinations of any two words in the code sequences. The goal of this supervised training is to propagate the error backwards and update the parameters of the linear mapping matrix $B$.

\subsection{Syntax Tree Induction}\label{sec_grammar_induction}
\begin{algorithm}[t!]
	\SetKwInOut{Input}{Require}\SetKwInOut{Output}{output}
    $S=(w_1,w_2,\ldots,w_n)$: a sequence of code tokens, with length $n$\;
    $\mathbf{d}=(d_1,d_2,\ldots,d_{n-1})$: a vector whose elements are the distance between two adjacent code tokens\;
    \SetKwFunction{FMain}{Tree}
    \SetKwProg{Fn}{Function}{:}{}
	\Fn{\FMain{$S, \mathbf{d}$}}{
        \eIf{$ \mathbf{d} = [] $}
            {$ node \gets Leaf(S_0);$}
            {
                $ i = \operatorname{argmax}(\mathbf{d})$\;
                $child_l = \textsf{Tree}(S_{\leq i},\mathbf{d}_{<i})$\;
                $child_r = \textsf{Tree}(S_{>i},\mathbf{d}_{>i})$\;
                $node \gets Node(child_l, child_r)$\;
            }
        \textbf{return} $ node; $ 
    }
    \textbf{End Function}
	\caption{Greedy top-down binary syntax tree induction based on syntax distances.}
	\label{alg_distance_to_tree}
\end{algorithm}
In this approach, we propose to investigate the capability of pre-trained code model in inducing syntax structure, \textit{without} training.
The key insight of our approach is that  
if the distance between two tokens is close (e.g., with a similar attention distribution, or with a similar representation),
they are expected to be close in the syntax tree, i.e., sharing the same parent.
Based on this insight, we propose to induce the syntax tree from the distances between two tokens. 
Our assumption is that if the induced tree derived from the pre-training model is similar to the gold standard syntax tree, we can reasonably infer that the syntactic structures have been preserved during the model pre-training.

We propose to induce the syntax tree based on the syntactic distance among code tokens, which was first introduced for grammar induction for natural languages~\cite{shen2018straight}.
Formally, given a code sequence $(w_1,w_2,\ldots,w_n)$, 
we compute $\mathbf{d}=(d_1,d_2,\ldots,d_{n-1})$, where $d_i$ corresponds to the syntactic distance between tokens $w_i$ and $w_{i+1}$. 
Each $d_i$ is defined as follows:
\begin{equation}
    d_i=f(g(w_i), g(w_i+1))\,,
\end{equation}
where $f(\cdot,\cdot)$ and $g(\cdot)$ denote the function of distance measurement and code representation learning, respectively. 
Here, we measure the syntactic distance between two tokens from their intermediate representation vector, as well as the self-attention distribution, with various distance measurement functions.
Specifically, let $g_l^v$ and $g_{l,k}^d$ denote the functions to generate the intermediate representation and self-attention in $l$-th layer and $k$-th head. 
To calculate the similarity between vectors, we have many options, in terms of the intermediate representation and attention distributions.
For example, we can use $L1$ and $L2$ to calculate the similarity between two intermediate representation vectors.
We can use Jensen-Shannon divergence~\cite{manning1999foundations} and Hellinger distance~\cite{le2012asymptotics} to calculate the similarity between two attention distributions.
Table~\ref{tab:distance_measure} summarizes all the available distance measurement functions.

Once the distance vector $\mathbf{d}$ is computed, we can easily convert it to the target syntax tree through a simple greedy top-down inference algorithm based on recursive partitioning of the input, as shown in Algorithm~\ref{alg_distance_to_tree}. Alternatively, this tree reconstruction process can also be done in a bottom-up manner, which is left for further exploration~\cite{shen2018straight}.
\begin{table}[h!]
	\centering
	\caption{The definition of different functions of distance measurement to compute the syntactic distance between two adjacent words in a code sequence. Note that $r=g^v(w_i)$, $s=g^v(w_{i+1})$,  $P=g^d(w_i)$, $Q=g^d(w_{i+1})$, $h$ denotes the hidden embedding size, and $n$ denotes the length of code sequence.}
    \begin{tabular}{l|l}
        \hline
        \textbf{Function} & \textbf{Definition}\\
        \hline
        \multicolumn{2}{l}{Distance functions for intermediate representations}\\
        \hline
        $L1(r,s)$   & $\sum_{i=1}^{h}\left | r_i-s_i\right |$ \\
        	$L2(r,s)$   & $\sqrt{\sum_{i=1}^{h}\left (r_i-s_i \right )^2}$ \\ 
        \hline
        \multicolumn{2}{l}{Distance functions for attention distributions}\\
        \hline
        $JSD(P||Q)$ & $\sqrt{({D_{KL}\left ( P||M \right )+D_{KL}\left ( Q||M \right )})/{2}}$\\
        & where $\ M=(P+Q)/{2}$ \\
        & and $\ D_{KL}\left ( A||B \right )=\sum_{w\in c} A(w)\log\frac{A\left ( w \right )}{B\left ( w \right )}$ \\
        $HEL(P,Q)$  & $\frac{1}{\sqrt{2}}\sqrt{ \sum_{i=1}^{n}\left ( \sqrt{p_i}-\sqrt{q_i} \right )^2} $\\ 
        \hline
    \end{tabular}
	\label{tab:distance_measure}
\end{table}

\paragraph{Injecting Bias into Syntactic Distance}
From our observation, the AST of source code tends to be right-skewed.
This has also been a well-known bias in the constituency trees for English.
Therefore, it motivates us to influence the induced tree such that they are moderately right-skewed following the nature of gold-standard ASTs.
To achieve this goal, we propose to inject the inductive deviation into the framework by simply modifying the value of the syntactic distance. 
In particular, we introduce the right-skewness bias
to influence the spanning tree to make it right-biased appropriately~\cite{kim2020pre}. 
Formally, we compute $\hat{d}_i$ by appending the following linear bias term to every $d_i$:
\begin{equation}
    \hat{d}_i= d_i+\lambda\cdot AVG(\mathbf{d}) \times \left( 1-\frac{1}{(m-1)\times (i-1)} \right)\,,
\end{equation}
where AVG($\cdot$) outputs an average of all elements in a vector, $\lambda$ is a hyperparameter, 
and $i$ ranges from $1$ to $m$, where $m = n-1$.

Note that, introducing such a bias can also examine what changes are made to the resulting tree structure.
Our assumption is that: if injecting the bias does not affect the performance of the pre-trained model for unsupervised analysis, we can infer that they capture the bias to some extent.

\begin{figure}[h!]
	\centering
	\includegraphics[width=.46\textwidth]{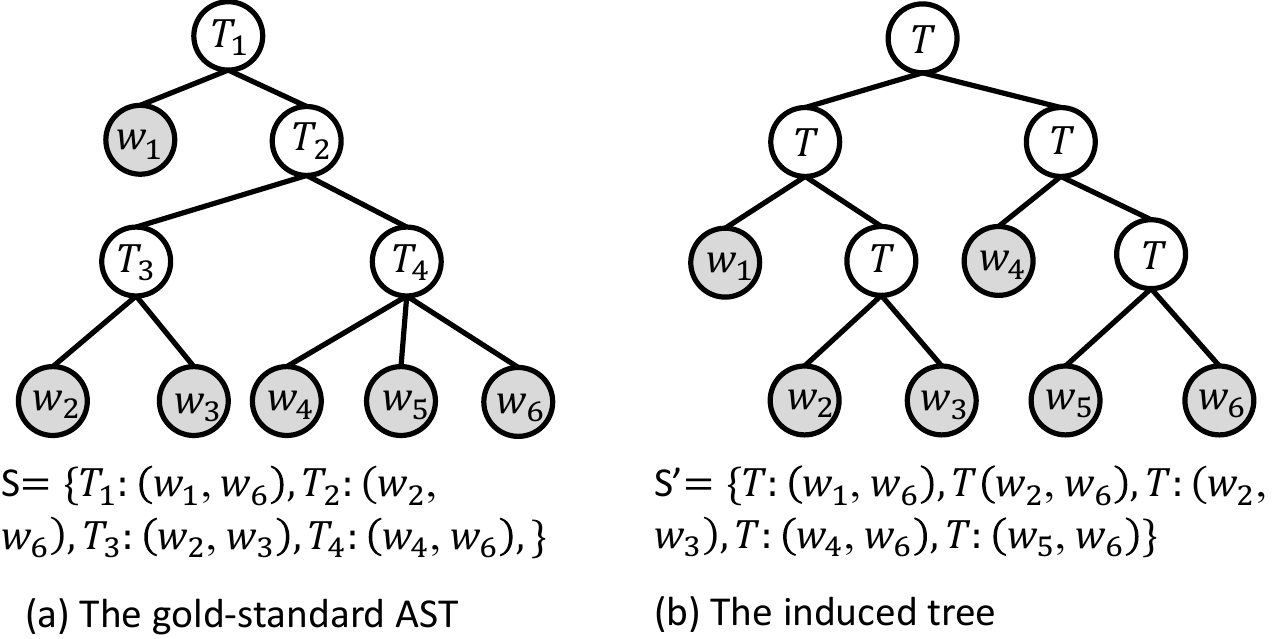}
	\caption{A toy example to illustrate the calculation of similarity between the gold-standard AST and induced tree.}
	\label{fig_tree_to_set}
	\vspace{-1em}
\end{figure}

\paragraph{Similarity between Two Trees}
Here we introduce the way we measure the similarity of the induced tree and the gold-standard AST.
Specifically, we first transform the tree structure into a collection of intermediate nodes, where each intermediate node is composed of two leaf nodes. Then we measure the similarity between the two collections.
Figure~\ref{fig_tree_to_set} shows a toy example to illustrate the calculation of similarity between two trees, i.e., the gold-standard AST (Figure~\ref{fig_tree_to_set}(a)) and induced tree (Figure~\ref{fig_tree_to_set}(b)).
As shown in Figure~\ref{fig_tree_to_set}(a), the gold-standard AST consists of four intermediate nodes (i.e., $T_1$, $T_2$, $T_3$, and $T_4$).
For each intermediate, we further expand it using two leaf nodes. For example, the $T_1$ is expanded into $(w_1, w_6)$, where $w_6$ is randomly selected from the $w_4$, $w_5$, and $w_6$.
Similarly, we also transform the induced tree into a collection of leaf nodes.
\begin{figure*}[t!]
	\centering
	\begin{subfigure}{0.32\textwidth}
		\centering
		\includegraphics[width=\textwidth]{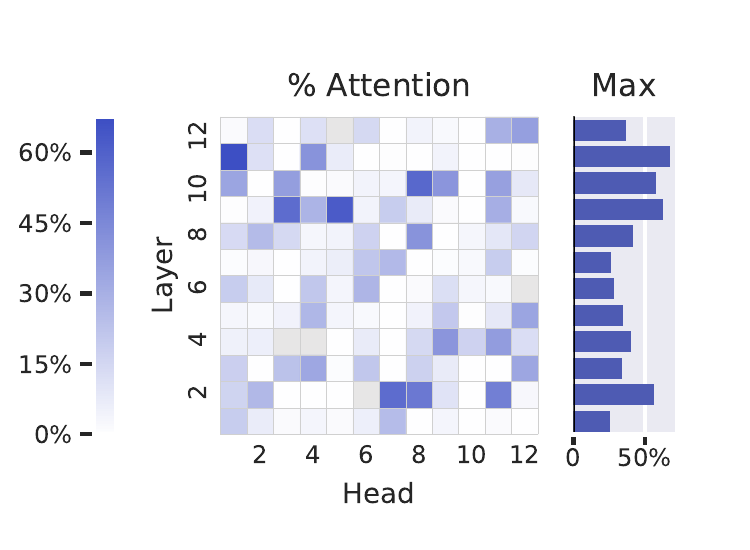}
		\caption{CodeBERT (Python)}
		\label{fig:codebert_python_contact_map}
	\end{subfigure}
	\begin{subfigure}{0.32\textwidth}
		\centering
		\includegraphics[width=\textwidth]{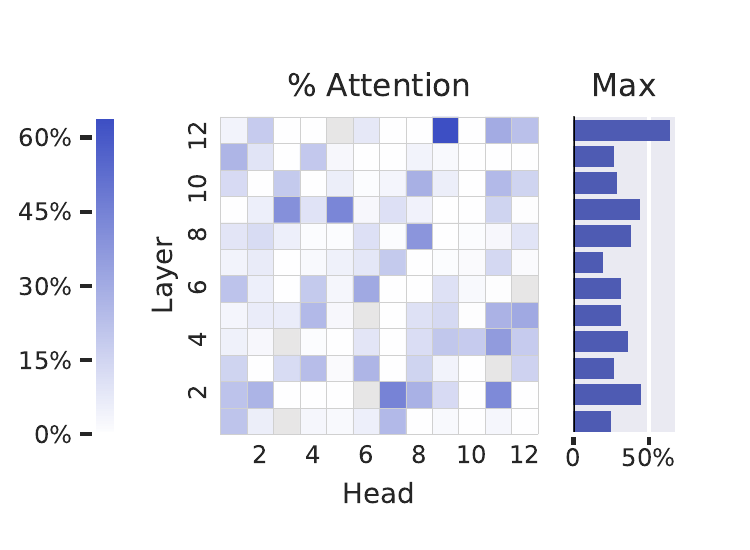}
		\caption{GraphCodeBERT (Python)}
		\label{fig:graphcodebert_python_contact_map}
	\end{subfigure}
	\begin{subfigure}{0.32\textwidth}
		\centering
		\includegraphics[width=\textwidth]{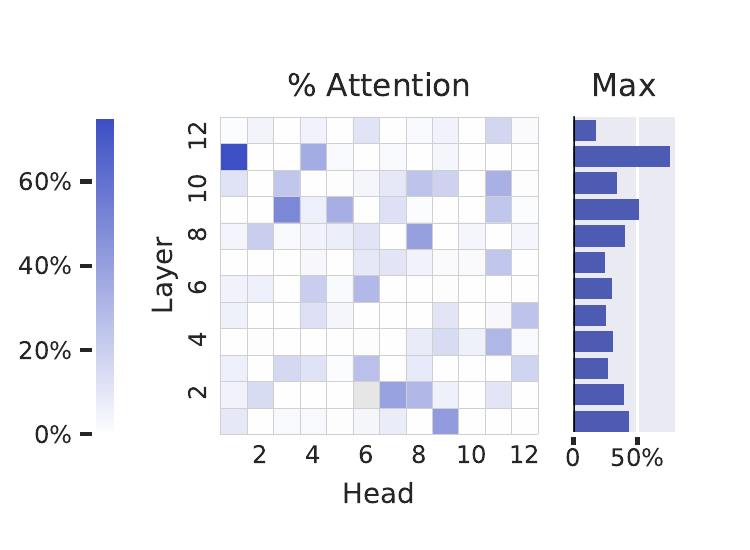}
		\caption{CodeBERT (Java)}
		\label{fig:codebert_java_contact_map}
	\end{subfigure}
	\begin{subfigure}{0.32\textwidth}
		\centering
		\includegraphics[width=\textwidth]{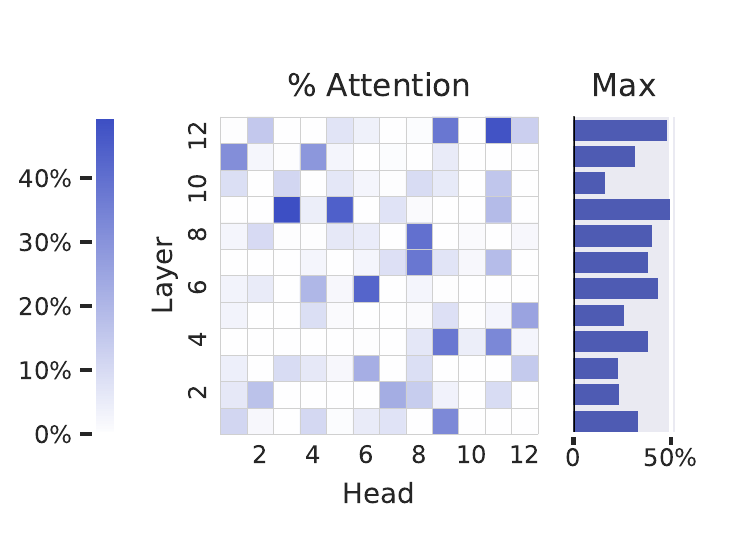}
		\caption{GraphCodeBERT (Java)}
		\label{fig:graphcodebert_java_contact_map}
	\end{subfigure}
	\begin{subfigure}{0.32\textwidth}
		\centering
		\includegraphics[width=\textwidth]{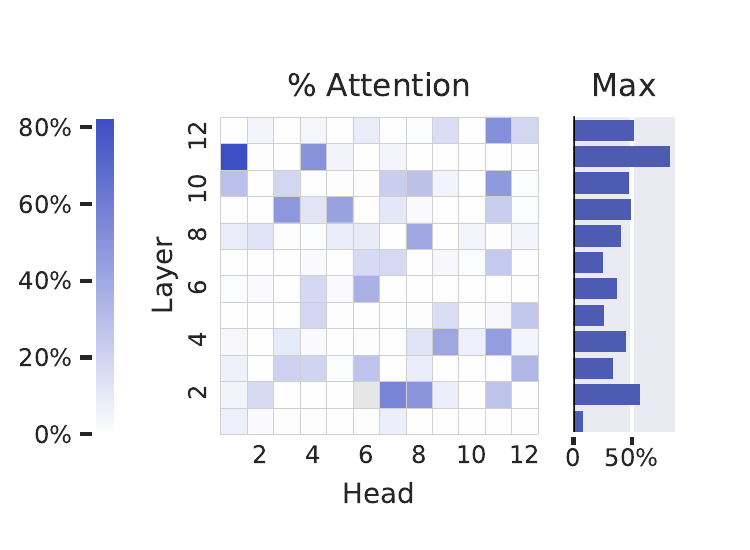}
		\caption{CodeBERT (PHP)}
		\label{fig:codebert_php_contact_map}
	\end{subfigure}
	\begin{subfigure}{0.32\textwidth}
		\centering
		\includegraphics[width=\textwidth]{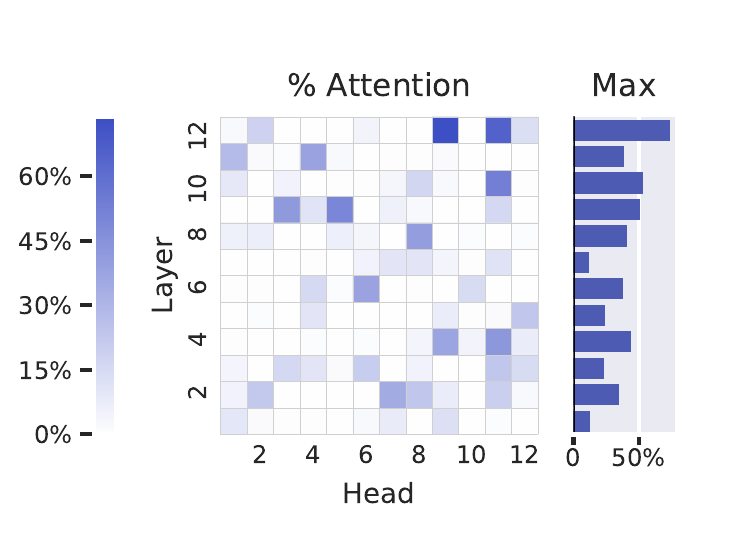}
		\caption{GraphCodeBERT (PHP)}
		\label{fig:graphcodebert_php_contact_map}
	\end{subfigure}
	\caption{Consistency between the attention and AST for the CodeBERT and GraphCodeBERT on different programming languages (i.e., Python, Java, and PHP). 
	These heatmaps show the proportion of high-confidence attention weights ($\alpha_{i,j} >\theta$) from each head which connect those code tokens that in the \textit{motif structure} of AST. The bars show the maximum value of each layer.
	}
    \label{fig:analysis_in_attention}
\vspace{-12pt}
\end{figure*}

Given two sets, we use the $F1$ score to measure their similarity.
Let $S$ denote the set of gold-standard tree, and $S'$ denote the set of predicted tree, 
we can calculate the precision and recall by
$precision = \frac{|S\cap S'|}{|S'|}$, and $recall = \frac{|S\cap S'|}{|S|}$, respectively.
The F1 score is the the harmonic mean of precision and recall, as follows:
\begin{equation}
    F1=2\ast\frac{precision \cdot recall}{precision + recall}\,.
\end{equation} 


\section{Experimental Design and Results}\label{sec_experiment}
In this section, we conduct experiments to explore what the pre-trained code models capture from three distinct aspects, i.e., attention analysis, structural probing on word embedding, and syntax tree induction.
\subsection{Experimental Design}
We investigate two Transformer-based pre-trained models (i.e., CodeBERT~\cite{feng2020codebert} and GraphCodeBERT~\cite{guo2020graphcodebert}), both of which are composed of 12 layers of Transformer with 12 attention heads.
These models are both
pre-trained on CodeSearchNet~\cite{husain2019codesearchnet}, a large-scale of code corpora collected from GitHub across six programming languages. 
The size of representation in each Transformer layer is set to 768.
Without loss of generability, we select Python, Java, and PHP as our target programming languages and use the corresponding dataset from CodeSearchNet. 
For all experiments, we exclude the attention to the \texttt{[SEP]} delimiter as it has been proven to be a ``no-op'' attention mark~\cite{clark2019does}, as well as the attention to the \texttt{[CLS]} mark, which is not explicitly used for language modeling.
Note that, in the pre-training phase, the input code snippets have been tokenized into subwords via \textit{byte-pair encoding} (BPE)~\cite{sennrich2015neural} before being passed to the pre-trained model.
However, our analyses are all based on the word-level code tokens. Therefore, we represent each word by averaging the representations of its subwords.
All the experiments were conducted on a Linux server, with 128GB memory, and a single 32GB Tesla V100 GPU.

Through comprehensive analysis, we aim to answer the following research questions:
\begin{itemize}
    \item \textbf{RQ1 (Attention Analysis):} Does attention align with the syntax structure in source code?
    \item \textbf{RQ2 (Probing on Word Embedding):} Are the syntax structure encoded in the contextual code embedding? 
    \item \textbf{RQ3 (Syntax Tree Induction):} Are the pre-trained code model able to induce the syntax structure of code?
\end{itemize}

\begin{figure*}[t]
	\centering
	\includegraphics[width=0.96\textwidth]{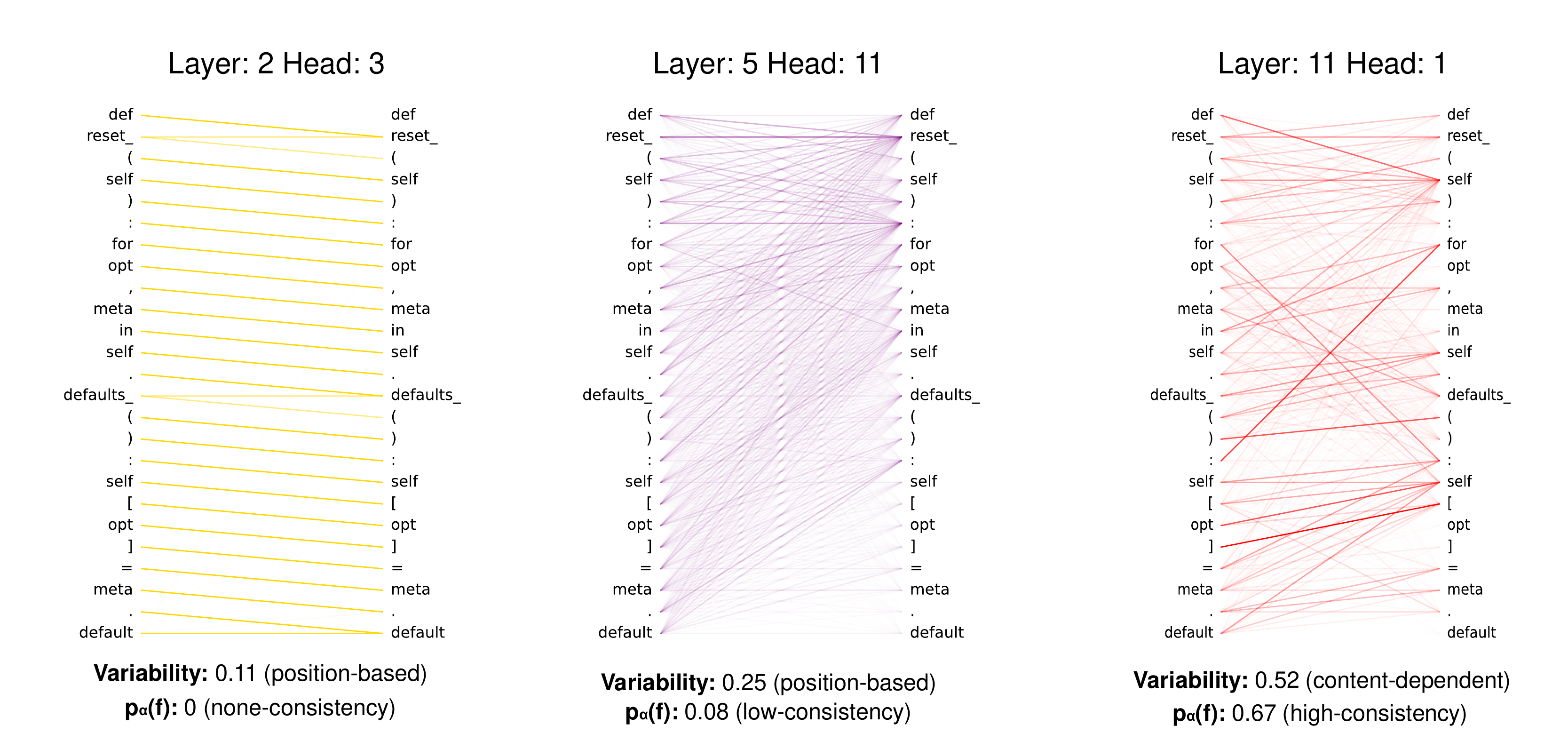}
	\caption{
	Visualization of attention heads in CodeBERT, along with the value of attention analysis ($p_{\alpha}(f)$), and attention variability, given a Python code snippet.
	Left: Attention visualised in Layer 2, Head 3, which focuses attention primarily on the position of next token. Center: Attention visualized in Layer 5, Head 11, which disperses attention roughly evenly across all tokens. 
		Right:  Attention visualized in Layer 11, Head 1, which focuses on the content, and is highly aligned with the AST. 
	}
	\label{fig:variability_example}
	\vspace{-1em}
\end{figure*}
\begin{figure}
	\centering
	\centering
	\includegraphics[width=0.44\textwidth]{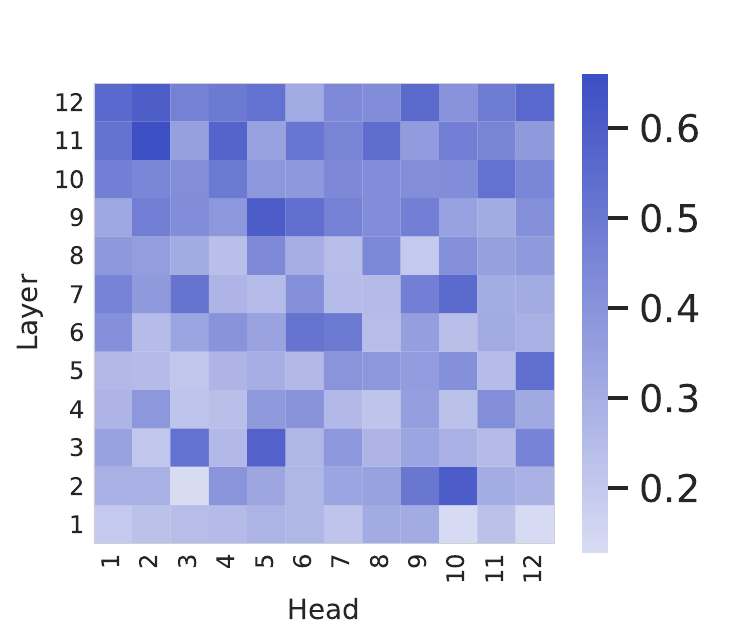}
	\caption{The variability of attention distribution by layer-head in Python. High-values indicate content-dependent heads, and low-values indicate position-based heads.}
	\label{fig:variability}
	\vspace{-1em}
\end{figure}
\subsection{RQ1: Attention Analysis}
In attention analysis, we aim to investigate whether the attention aligns with the syntax structure of source code.
\paragraph{Experimental Settings}
Following~\cite{vig2020bertology}, we set the attention threshold $\theta$ in Equation~\ref{eq_attention_analysis} as $0.3$,
so as to ensure selecting high-confidence attention and  retaining enough data for our analysis. We leave the analysis of the impact of $\theta$ in future work.
Furthermore, our analysis is based on a subset of 5,000 code snippets randomly sampled from the training dataset.
In order to reduce the memory and accelerate the computation process, we truncate all the long code sequences within the length of $512$.
We only include the results of attention heads where 
at least 100 high-confidence attention scores are available in our analysis.

\paragraph{Experimental Results}
Figure~\ref{fig:analysis_in_attention} shows how attention aligns with the AST structure for CodeBERT and GraphCodeBERT on different programming languages (i.e., Python, Java, and PHP), according to the indicators defined in Equation~\ref{eq_attention_analysis}. 
The figure shows the proportion of high-confidence attention weights ($\alpha_{i,j} >\theta$) from each head
which connect those code tokens that in the \textit{motif structure} of AST.
The bar plots show the maximum score of each layer.
From this figure, we can observe that the most aligned heads are located in the deeper layers and the concentration is as high as $67.25\%$ (Layer 11, Head 1 in CodeBERT) and $59\%$ (Layer 12, Head 9 in GraphCodeBERT).
These high scores indicate that \textbf{attention aligns strongly with the \textit{motif structure} in AST, especially in the deeper layers.
} 
This is because the heads in deeper layers have stronger capabilities in capturing longer distances.


Although there is a strong alignment between the attention and the syntax structure of code, 
it is still necessary to distinguish whether the attention is based on the position or content of code token,
as mentioned in Sec.~\ref{sec_attention_analysis}.
In Figure~\ref{fig:variability_example}, we show the attention variability of attention heads in CodeBERT for a Python code snippet.
Figure~\ref{fig:variability_example} (left) and Figure~\ref{fig:variability_example} (center) show two examples of heads that put more focus on the position, respectively from Layer 5, Head 11, and Layer 2, Head 3.
Based on the variability defined in Equation~\ref{eq_variability}, we can see that the attention in Layer 5, Head 11, is evenly dispersed, with the variability of 0.25.
Moreover, in Layer 5, Head 11, it is apparent to see that the attention is focusing on the next token position.
Figure~\ref{fig:variability_example} (right) shows the content-dependent head from Layer 11, Head 1, which has the highest alignment relationship with the abstract syntax tree structure among all heads. 
In Figure~\ref{fig:variability}, we also visualize the variability of attention distribution by layer-head in Python. The high-values indicate the content-dependent heads, and the low-values indicate the position-based (or content-independent) heads.

\begin{tcolorbox}[left=2pt,right=2pt,top=0pt,bottom=0pt]
\textbf{Summary.} 
Through attention analysis, we find that the learned attention weights are strongly aligned with the \textit{motif structure} in an AST.
Additionally, each attention across different heads and layers put different focus on the position and content of the tokens of source code.
\end{tcolorbox}

\subsection{RQ2: Probing on Word Embedding}
We conduct structural probing on the word embedding of source code, to investigate whether the word embedding in the Transformer-based pre-trained model embeds the syntax structure of code.
\paragraph{Experimental Settings}
Given a pair of code tokens (leaf nodes) in AST, we measure the correlation between the predicted distance using word embedding and the gold-standard distance in the AST.
Specially, we use the Spearman correlation~\cite{corder2014nonparametric}
to measure the predicted distance vector and the gold-standard distance vector, among all samples of code snippets.
When training the linear transformation matrix $B$ in Equation~\ref{eq_mapping_matrix_learning}, we limit the code length to $100$.
We probe the contextual representations in each layer of Transformer, and denote the investigated pre-trained code models as CodeBERT-$K$ and GraphCodeBERT-$K$, where $K$ indexes the layer of Transformer in corresponding model. 
To serve as a comparison against the pre-trained code models, we also design a baseline model – CodeBERT-0, which denotes the simple word embedding before being fed into the Transformer layers.
In evaluation, we average the Spearman correlation between all fixed-length code sequences.  
We report the average value of the entire sequence length of $5\sim50$ as the Spearman metric, as in~\cite{hewitt2019structural}.
\begin{table}[t]
	\centering
	\caption{
The average Spearman correlation of probing in Python.
	}
\vspace{-1em}
	\begin{tabular}{lc}
		\hline
		\textbf{Method} & \textbf{Spearman Correlation} \\ 
		\hline
		CodeBERT-0 & 0.60 \\ 
		\hline
		CodeBERT-1 & 0.69 \\ 
		CodeBERT-5 & 0.85 \\ 
		GraphCodeBERT-5 & 0.86 \\ 
		\hline
	\end{tabular}
	\label{table_word_embedding_Dspr}
	\vspace{-1em}
\end{table}
\paragraph{Experimental Results}
Table~\ref{table_word_embedding_Dspr} shows the results of probing in Python. From this table,
we find that the CodeBERT-0 without Transformer layers achieves inferior performance than that with multiple layers of Transformer. This confirms our assumption that Transformer has the ability of capturing the syntax information of source code.
In addition, we can also find that GraphCodeBERT performs better than CodeBERT, indicting that it is helpful to explicitly incorporate the syntax structure into model pre-training.

\begin{figure}[t!]
	\centering
	\includegraphics[width=.4\textwidth]{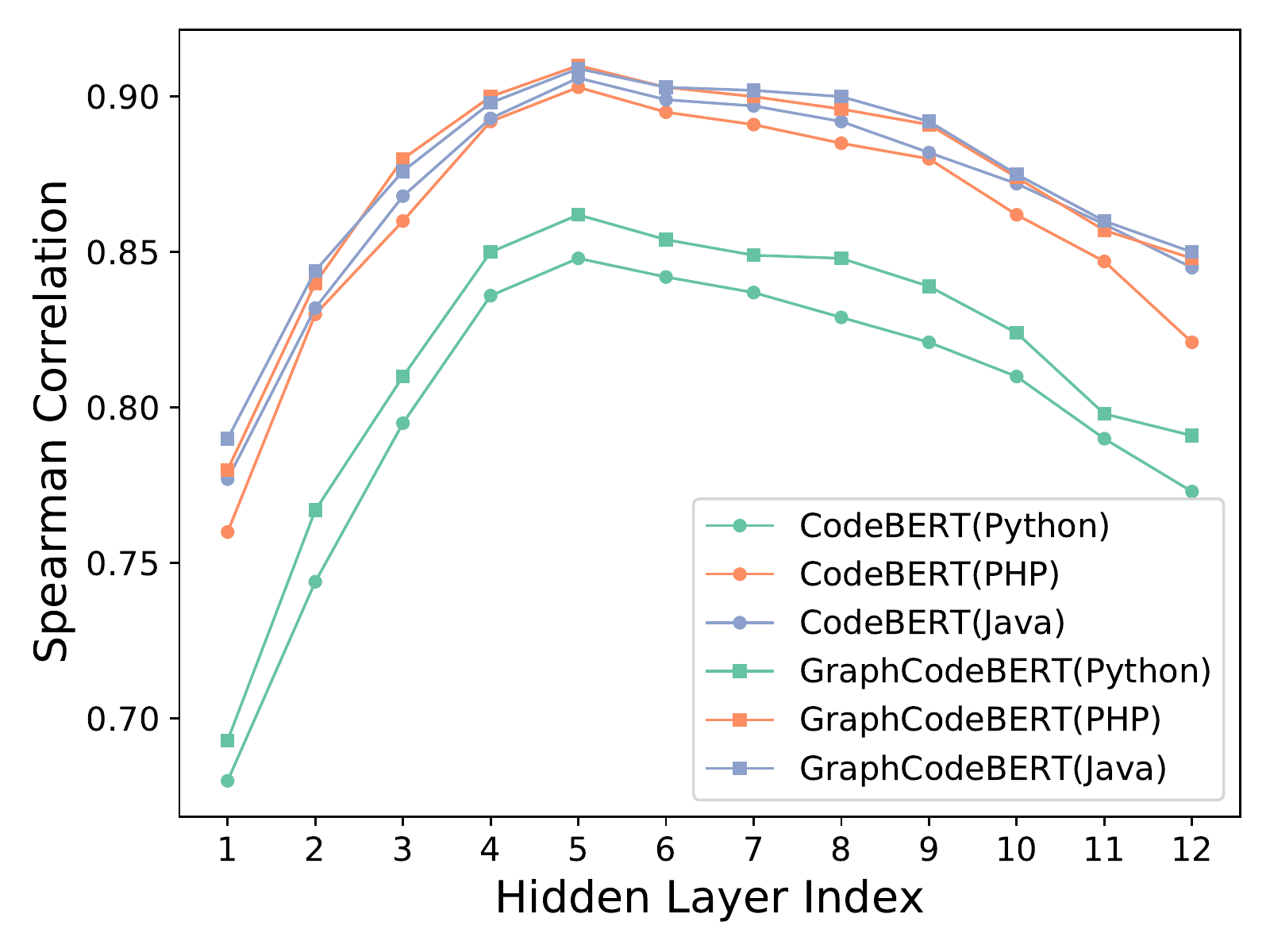}
	\caption{The average Spearman correlation for CodeBERT and GraphCodeBERT in multiple programming languages.
	}
	\label{fig:DSpr in model layers}
	\vspace{-1em}
\end{figure}

Figure~\ref{fig:DSpr in model layers}
shows the Spearman correlation of probing on the representation in each layer of the models. 
We can observe that \textbf{the capability of capturing syntax structure differs across different layers of Transformer}. The best performance is obtained in the 5-th layer.
For example, in Python, CodeBERT and GraphCodeBERT achieve the highest Spearman correlation (84\% and 86\%, respectively) in the 5-th layer. %
Furthermore, in each layer of Transformer, GraphCodeBERT still performs better than CodeBERT in capturing the syntax structure of programs written Python, Java, and PHP, confirming the observation from Table~\ref{table_word_embedding_Dspr}.

\begin{figure*}[t]
    \begin{subfigure}{0.24\textwidth}
        \centering
        \includegraphics[width=\textwidth]{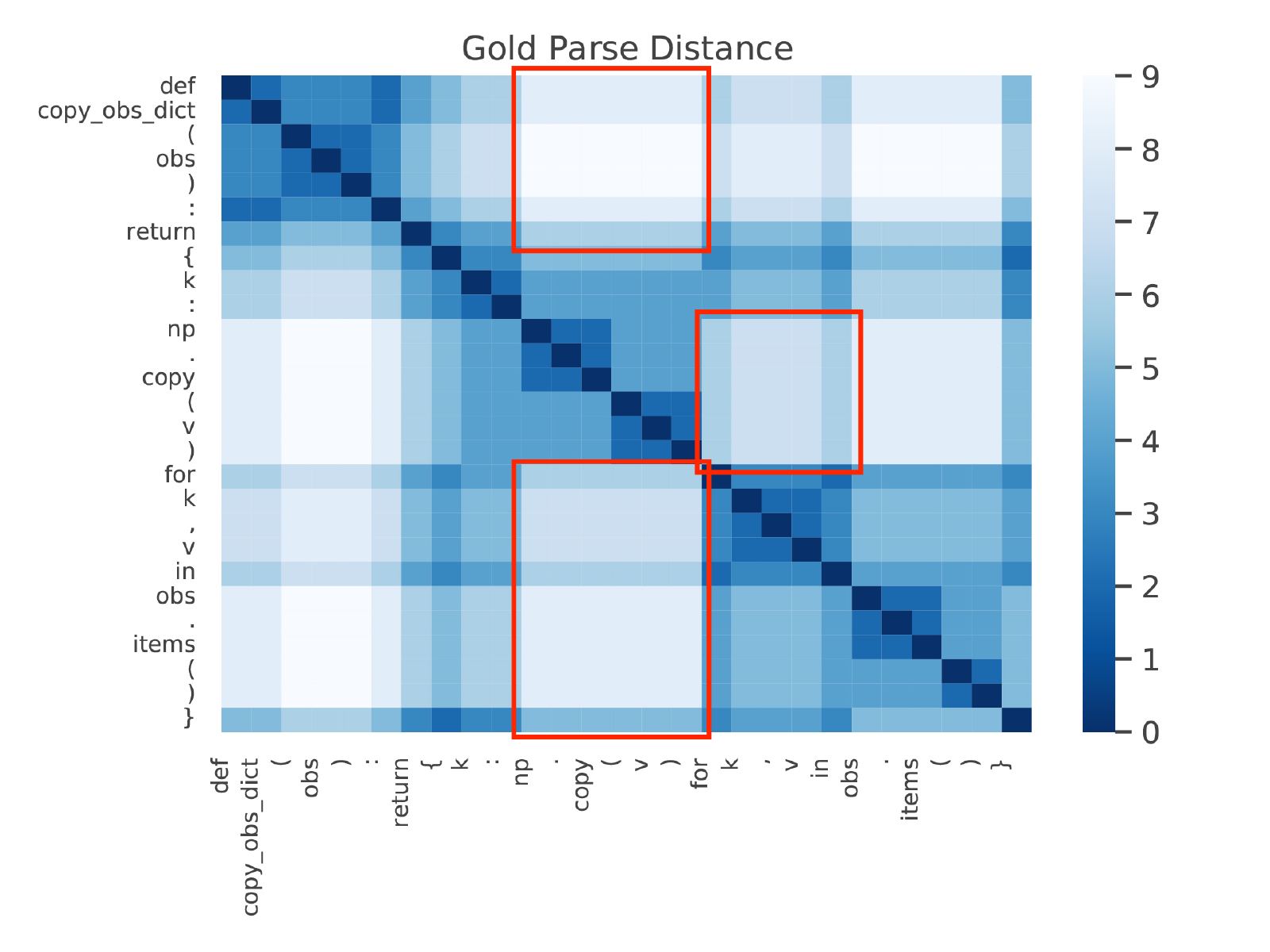}
        \caption{Gold-standard}
        \label{fig:word_embedding_predict_distance_a}
    \end{subfigure}
    \begin{subfigure}{0.24\textwidth}
        \centering
        \includegraphics[width=\textwidth]{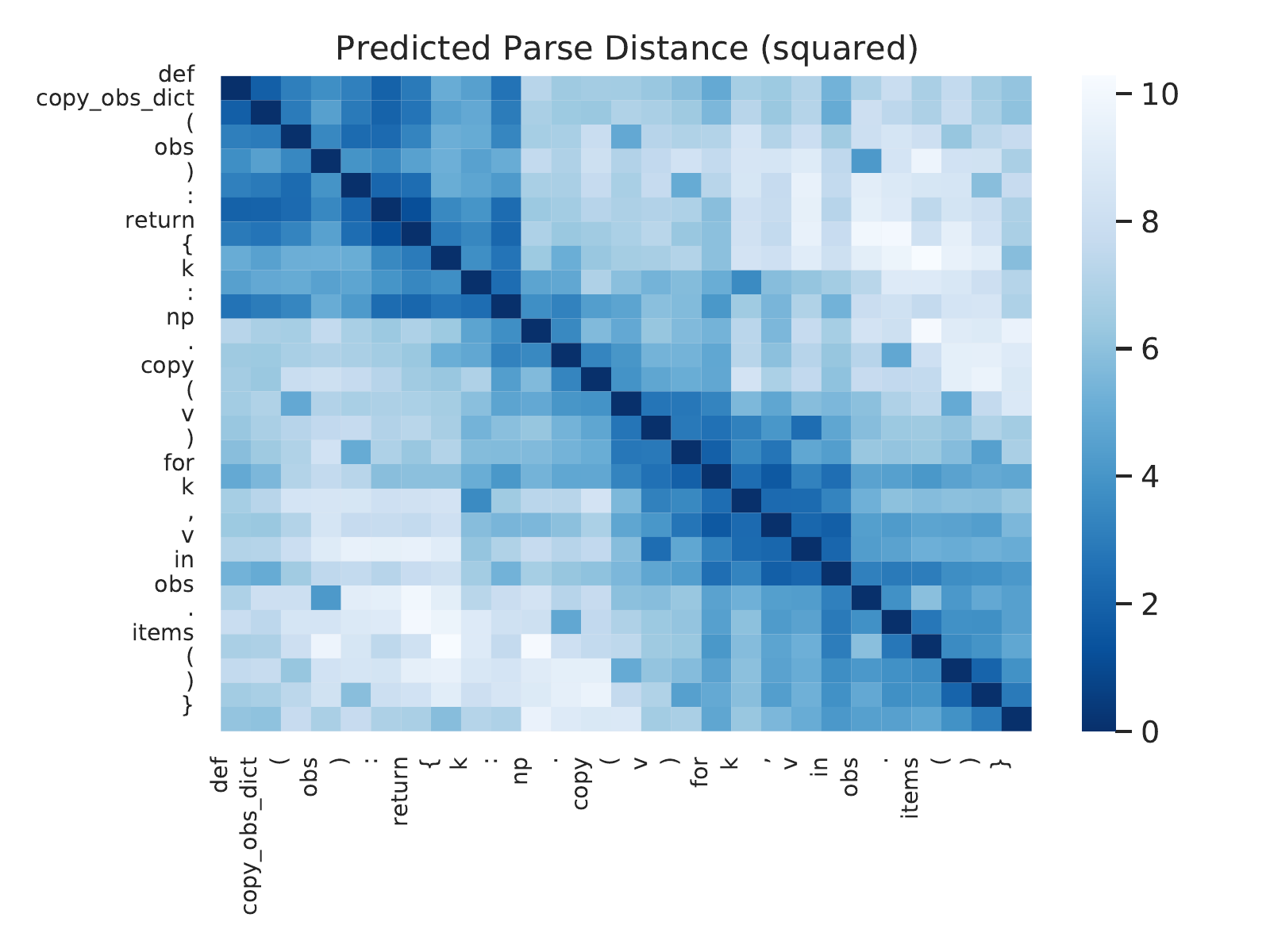}
        \caption{CodeBERT-0 prediction}
        \label{fig:word_embedding_predict_distance_b}
    \end{subfigure}
    \begin{subfigure}{0.24\textwidth}
        \centering
        \includegraphics[width=\textwidth]{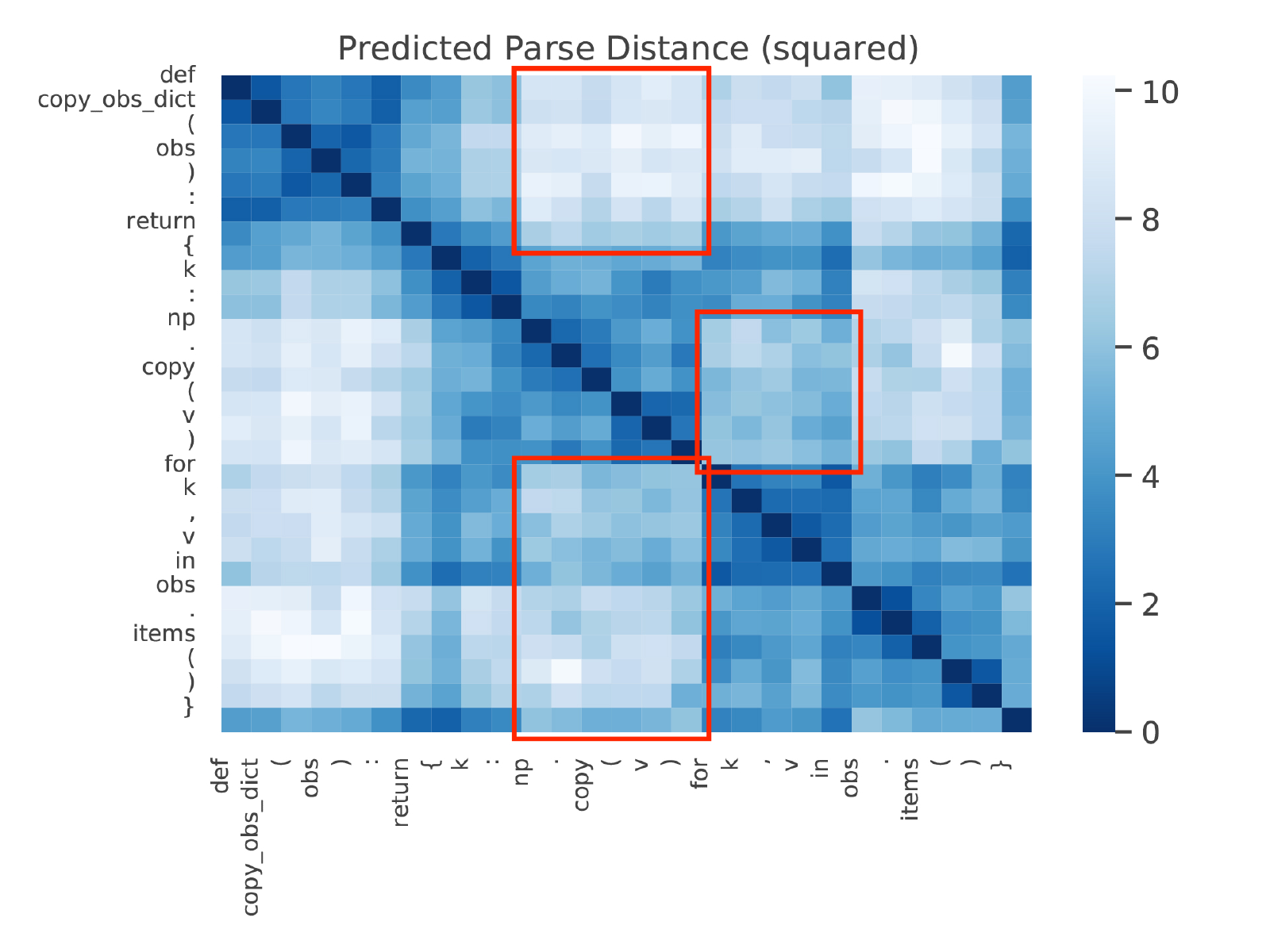}
        \caption{CodeBERT-5 prediction}
        \label{fig:word_embedding_predict_distance_c}
    \end{subfigure}
    \begin{subfigure}{0.24\textwidth}
        \centering
        \includegraphics[width=\textwidth]{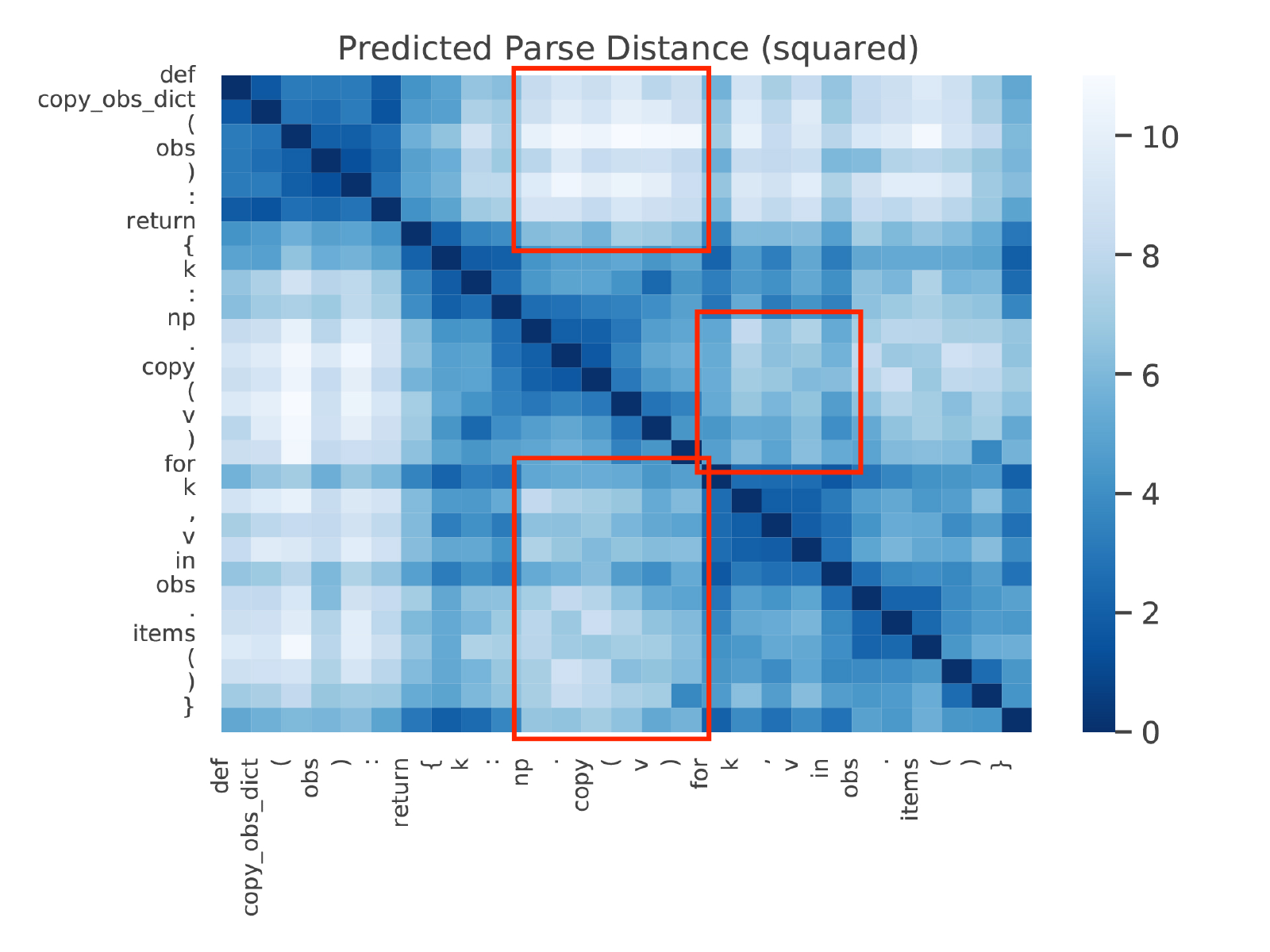}
        \caption{GraphCodeBERT-5 prediction}
        \label{fig:word_embedding_predict_distance_d}
    \end{subfigure}
    \caption{
    The heatmaps of gold-standard distance and predicted distance based on pre-trained CodeBERT and GraphCodeBERT models in Python.
    (a) Gold-standard tree distance between all pairs of code tokens in AST.
    (b-d) The predicted distance based on the probing of CodeBERT and GraphCodeBERT. 
    The darkness of color indicates the closeness of the paired words.
    }
    \label{fig:word_embedding_predict_distance}
\end{figure*}

Figure~\ref{fig:word_embedding_predict_distance} 
shows the heatmaps of gold-standard and predicted distances based on pre-trained CodeBERT and GraphCodeBERT, for a given input Python code snippet.
We can see that Figures~\ref{fig:word_embedding_predict_distance_c} and \ref{fig:word_embedding_predict_distance_d} look more similar to the gold-standard one (Figure~\ref{fig:word_embedding_predict_distance_a}) than to Figure~\ref{fig:word_embedding_predict_distance_b}.
In these figures, some matching parts are marked in red.
The result confirms that CodeBERT-5 and GraphCodeBERT-5 (with multiple layers of Transformer) perform better than CodeBERT-0 (without passing through the Transformer layers).

\begin{tcolorbox}[left=2pt,right=2pt,top=0pt,bottom=0pt]
\textbf{Summary.} Through embedding analysis, we can observe that the syntax structure of code has been well preserved in different hidden layers of the pre-trained language models (i.e., CodeBERT and GraphCodeBERT).
\end{tcolorbox}
\begin{table*}[t!]
	\centering
	\caption{Results of syntax tree induction in Python. f: function of distance measurement, L: layer number, A: attention head number, AVG: the average of all attentions.}
	\begin{tabular}{lcccccccccc}
		\hline
		\textbf{Model} & \textbf{f} & \textbf{L} & \textbf{A} & \multicolumn{1}{l}{\textbf{F1}} & \multicolumn{1}{l}{\textbf{\texttt{parameters}}} & \multicolumn{1}{l}{\textbf{\texttt{attribute}}} & \multicolumn{1}{l}{\textbf{\texttt{argument}}} & \multicolumn{1}{l}{\textbf{\texttt{list}}} & \multicolumn{1}{l}{\textbf{\texttt{assignment}}} & \multicolumn{1}{l}{\textbf{\texttt{statement}}} \\ 
		\hline
		\multicolumn{11}{l}{\textbf{Baselines}} \\ \hline
		Random Trees & - & - & - & 16.93 & 20.26\% & 30.75\% & 32.15\% & 24.34\% & 12.98\% & 16.03\% \\
		Balanced Trees & - & - & - & 16.79 & 0.46\% & 32.60\% & 30.85\% & 25.54\% & 14.25\% & 15.90\% \\
		Left Branching Trees & - & - & - & 18.49 & 23.25\% & 43.26\% & 50.99\% & 26.77\% & 5.78\% & 14.48\% \\
		Right Branching Trees & - & - & - & 26.36 & 44.07\% & 43.19\% & 37.86\% & 34.18\% & 8.34\% & 22.74\% \\
		CodeBERT-0 & - & - & - & 19.13 & 11.67\% & 25.54\% & 53.85\% & 27.62\% & 18.68\% & 21.89\% \\ \hline
		\multicolumn{11}{l}{\textbf{Pre-Trained Models (w/o bias)}} \\ \hline
		CodeBERT & JSD & 8 & AVG & 45.37 & 40.99\% & 66.65\% & 88.42\% & 56.90\% & 70.47\% & 66.10\% \\
		GraphCodeBERT & HEL & 8 & 10 & 51.34 & 95.96\% & 75.50\% & 67.76\% & 80.87\% & 72.88\% & 63.98\% \\ \hline
		\multicolumn{11}{l}{\textbf{Pre-Trained Models (w/bias $\lambda=1$)}} \\ \hline
		CodeBERT & HEL & 9 & AVG & 50.18 & 67.37\% & 67.93\% & 76.84\% & 71.60\% & 56.12\% & 62.93\% \\
		GraphCodeBERT & HEL & 9 & AVG & 54.80 & 74.68\% & 72.05\% & 84.18\% & 73.68\% & 76.10\% & 72.69\% \\ \hline
	\end{tabular}
	\label{syntax induction table}
	\vspace{-1em}
\end{table*}
\subsection{RQ3: Syntax Tree Induction}
We investigate the extent to which pre-trained code models capture the syntax structure of code by inducing a tree.
\paragraph{Experimental Settings}

For comparison, we introduce four traditional greedy top-down tree induction baselines for comparison, e.g., random, balanced, left-branching, and right-branching binary trees.
Take the random-based approach as an example, we recursively partition the code snippet based on a randomly selected position.
In addition, we also derive another baseline, CodeBERT-0, which is based on the word embedding before being fed into Transformer layers.
When injecting bias into the syntactic distance, 
we set the hyperparameter of bias $\lambda$ to $1$.
Due to the space limitation, we only report the F1 scores for six common intermediate nodes in Python AST, i.e., \texttt{Parameters}, \texttt{Attribute}, \texttt{Argument}, \texttt{List}, \texttt{Assignment}, and \texttt{Statement}.

\textit{Experimental Results.}
Table~\ref{syntax induction table} presents the results of various models for syntax tree induction on the test dataset. 
From this table, we can observe that \textbf{the right-branching tree induction approach achieves the best performance among all the baselines, confirming our assumption that the AST tends to be right-skewed.} 
When comparing the pre-trained code models (i.e., CodeBERT and GraphCodeBERT) with other baselines, it is clear to see the pre-trained code models significantly outperform other baselines, even without bias injection.
These results show that the \textbf{Transformer-based pre-training models are more capable of capturing the syntax structure of code to a certain extent through pre-training on a large-scale code corpus.} 
When comparing the pre-trained models w/ and w/o bias injection, we can observe that injecting bias can increase the performance of syntax tree induction up to 5\%.
This improvement indirectly shows that the current pre-trained code models do not capture well the property of the right-skewness of  AST.
It is worthy mentioning that the performance of \texttt{assignment} has been reduced after injection the bias. 
One possible hypothesis is that although the AST shows a right-skewness trend as whole, 
several subtrees (e.g., the subtree of \texttt{assignment}) are not right-skewed.  

Note that, in the experiments, we have tried different distance functions (as shown in Table~\ref{tab:distance_measure}) to measure distance based on attention and contextual representation in each Transformer layer.
Due to the space limitation, in Table~\ref{syntax induction table}, we only present the best performance when using different distance functions for each Transformer layer.
We can find that the JSD and HEL distance functions that are performed over attention distributions perform better than those over contextual word representations.
It shows that parsing trees from attention information is more effective than extracting from the contextual representation of pre-trained code models.

\begin{figure*}[t!]
	\centering
	\includegraphics[width=.98\textwidth]{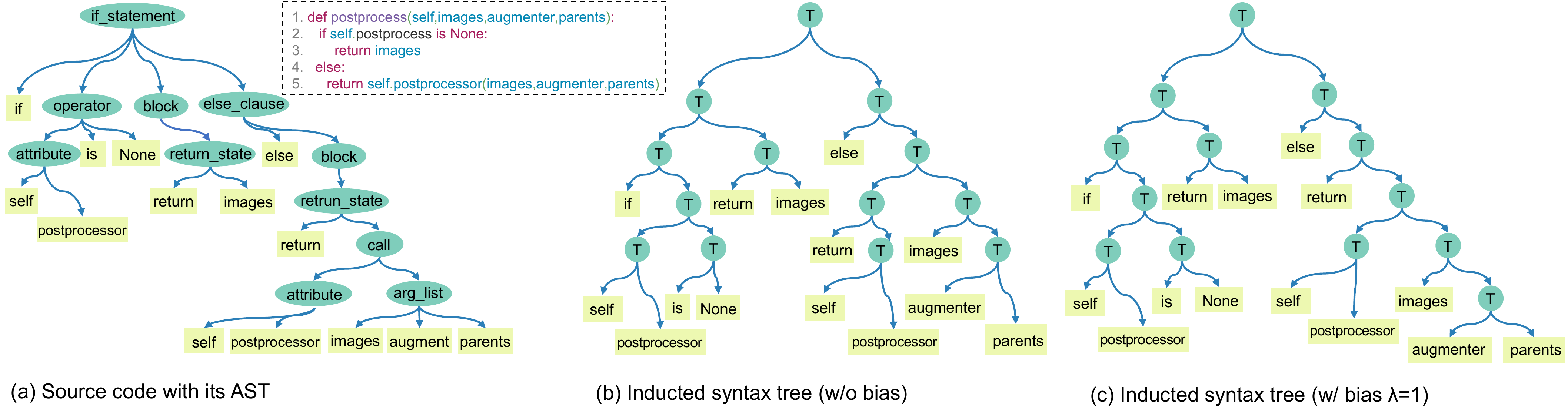}
    \caption{
    A case study of syntax tree induction based on CodeBERT for a given Python code snippet.
   }
	\label{fig:predict_tree}
	\vspace{-1em}
\end{figure*}

In Figure~\ref{fig:predict_tree}, we also show a case study of code snippet, with the induced trees based on CodeBERT, with and without bias injected. From this figure, we can see that several motif structures have been captured by CodeBERT, e.g., \texttt{return-images}, and \texttt{self-postprocessor}. It verifies the effectiveness of the syntax tree induction.

\begin{tcolorbox}[left=2pt,right=2pt,top=0pt,bottom=0pt]
\textbf{Summary.} 
The syntax tree of code can be induced by the pre-trained language models for code, to some extent.
In addition, extracting parse trees from attention information is more effective than extracting from the contextual representations of pre-trained code models.
\end{tcolorbox}

\section{Discussion}\label{sec_discussion}
\subsection{Observed Findings}
Through comprehensive analysis from three 
perspectives 
of pre-trained code models, we observe several insightful findings, which may inspire future study.
From attention analysis, we find that a word's attention distribution can align with the AST. The attentions align better with syntax connection in deeper layers than lower layers in the self-attention network. 
Moreover, we find that there exit position-based heads, which do not consider the context of text. It could suggest that if we remove these heads, it will not affect the final results and we can reduce the number of parameters of the pre-trained models. 
Then, we find the pre-trained models can embed syntactic information in the hidden layers. All pairs of words know their syntactic distance, and this information is a global structural property of the vector space. 
Finally, we use a simple tree construction algorithm to induce a syntax tree from pre-trained models. The results indicate that the pre-trained model such as CodeBERT is capable of perceiving the syntactic information to a certain extent when training on a large corpus. 
Our findings suggest that grammatical information can be learned by the pre-trained model, which could explain why a pre-trained model such as CodeBERT can achieve promising results in a variety of source code related downstream tasks such as code summarization, code search, and clone detection.


\subsection{Limitations and Future Work}
One limitation of our work is that the adopted structural analysis approaches are based on the AST structure of code, which could be just one aspect where the pre-trained models can achieve good results for source code. 
In our future work, we will investigate how the pre-trained models learn other aspects of source code, such as code tokens, {control-flow graphs} (CFGs), and {data-flow graphs} (DFGs).
Besides, in this paper, we only investigate two representative self-supervised pre-trained language models for source code, i.e., CodeBERT and GraphCodeBERT. It will be interesting to extent the analysis to other supervised learning models, as well as other deep neural networks (e.g., LSTM and CNN).

With regard to the design of the structural analysis approaches adopted in this paper, one limitation is that the structural probing on word embedding we currently use is relative simple. It would be interesting to develop a deep neural network to learn better mapping functions.
Meanwhile, the tree construction algorithm we use is a relatively simple top-down recursive binary tree algorithm. The \textit{right-skewness} bias we use was originally designed for the constituency trees in natural languages (e.g., English), which can be improved for ASTs.
Lastly, the AST structure is more complex than the induced tree, therefore there is still ampler room for improvement in the grammar induction algorithm. 
\section{Related Work}
Recently, there has been much effort in interpreting the BERT models in the NLP community.
At a high level, these interpretation approaches are developed from two perspectives: (1) interpreting the learned embedding, and (2) investigate whether BERT can learn  syntax and semantic information of natural languages. 
To interpret the learned embedding, 
\citet{ethayarajh2019contextual} studies 
whether the contextual information are preserved in the word embedding learned from pre-training models, 
including BERT, ELMo, and GPT-2. 
\citet{mickus2019you} systematically evaluate the pre-trained BERT using a distributed semantics models.
\citet{conneau2018you} and \citet{liu2019linguistic} design several probing tasks to investigate whether the sentence embedding can capture the linguistic properties.

To investigate the syntax and semantic knowledge in BERT,
\citet{tenney2019bert} develop a series of edge probing tasks to explore how the syntactic and semantic structure  can be extracted from  different layers of  pre-trained BERT.
\citet{htut2019attention} propose to extract implicit dependency relations from the attention weights of each layer/head through two approaches: taking the maximum attention weight and computing the maximum spanning tree.
\citet{hewitt2019structural} propose a structural probing approach to investigate whether the syntax information are preserved in word representations.

Specially, there also exists another line of work on visualizing attentions to investigate  which part of the feature space the model put more focus.
\citet{kovaleva2019revealing} study self-attention and conduct a qualitative and quantitative analysis of the information encoded by individual BERT's heads.
\citet{hoover2020exbert} introduce a tool, called exBERT, to help humans conduct flexible, interactive investigations and formulate hypotheses during the model-internal reasoning process.
Following this line of research, this paper proposes to extend and adapt the interpretation techniques from the NLP community to understand and explain what feature correlations can be captured by a pre-trained code model in the embedding space.

\section{Conclusion}\label{sec_conclusion}
In this paper, we have explored the interpretability of pre-trained language models for source code  (e.g., CodeBERT, GraphCodeBERT). We conduct a thorough structural analysis from the following three aspects, aiming to give an interpretation of pre-trained code models. First, 
we analyze the self-attention weights and align the weights with the syntax structure.
Second, we propose a structural probing approach to investigate whether the contextual representations in Transformer capture the syntax structure of code.
Third, we investigate whether the pre-trained code models have the capability of inducing the syntax tree without training.
The analysis in this paper has revealed several interesting findings that can inspire future studies on code representation learning.

\paragraph{Artifacts.} 
All the experimental data and source code used in this work will be integrated into the open-source toolkit \textsc{NaturalCC}~\cite{wan2022naturalcc}, which is available at \url{https://github.com/CGCL-codes/naturalcc}.

\begin{acks}
This work is supported by National Natural Science Foundation of China under grand No. 62102157. This work is also partially sponsored by Tencent Rhino-Bird Focus Research Program of Basic Platform Technology.
We would like to thank all the anonymous reviewers for their constructive comments on improving this paper.
\end{acks}
\balance
\bibliographystyle{ACM-Reference-Format}
\bibliography{ref}
\end{document}